\documentclass[preprint,12pt,authoryear,times]{elsarticle}
\usepackage[nolists,heads,nomarkers]{endfloat}
\usepackage{graphicx}
\usepackage{amsmath,amssymb}
\usepackage{bm}
\newcommand{\tr}{^{\sf T}}
\newcommand{\mom}[1]{\ensuremath{\langle #1 \rangle}}

\usepackage{psfrag}
\usepackage[dvipsnames,rgb,dvips]{xcolor}
\graphicspath{{./}}
\newcommand{\ve}[1]{\bm{#1}}
\newcommand{\ven}[1]{\bm{\mathit{#1}}}
\newcommand{\ma}[1]{\mathbf{#1}}
\newcommand{\dd}{{\rm d}}
\usepackage[normalem]{ulem}

\newcommand{\nn}{\nonumber}

\newcommand{\ket}[1]{\ve #1}
\usepackage[title]{appendix}
\usepackage{booktabs}
\usepackage{longtable}
\usepackage[T1]{fontenc}
\journal{Theoretical Population Biology}
\begin{document}

\begin{frontmatter}
\title{Metapopulation dynamics on the brink of extinction}

\author[cam]{A. Eriksson}
\author[ugo]{F. {El\'ias-Wolff}\;\!} 
\author[ugo]{B. Mehlig}
\address[cam]{Department of Zoology, University of Cambridge, Cambridge, CB2 3EJ, UK}
\address[ugo]{Department of Physics, University of Gothenburg, SE-41296 Gothenburg, Sweden}

\begin{abstract}
We analyse metapopulation dynamics in terms of an individual-based, stochastic model of a finite metapopulation. We suggest a new approach, using the number of patches in the population as a large parameter. This approach does not require that the number of individuals per patch is large, neither is it necessary to assume a time-scale separation between local population dynamics and migration. Our approach makes it possible to accurately describe the dynamics of metapopulations consisting of many small patches. We focus on metapopulations on the brink of extinction. We estimate the time to extinction and describe the most likely path to extinction. We find that the logarithm of the time to extinction is proportional to the product of two vectors, 
a vector characterising the distribution of patch population sizes in the quasi-steady state, and a vector -- related to Fisher's reproduction vector -- that quantifies the sensitivity of the quasi-steady state distribution to demographic fluctuations. We compare our analytical results to stochastic simulations of the model, and discuss the range of validity of the analytical expressions. By identifying fast and slow degrees of freedom in the metapopulation dynamics, we show that the dynamics of large metapopulations close to extinction is approximately described by a deterministic equation originally proposed by Levins (1969). We were able to compute the rates in Levins' equation in terms of the parameters of our stochastic, individual-based model. It turns out, however, that the interpretation of the dynamical variable depends strongly on the intrinsic growth rate and carrying capacity of the patches. Only when the growth rate and the carrying capacity are large does the slow variable correspond to the number of patches, as envisaged by Levins. Last but not least, we discuss how our findings relate to other, widely used metapopulation models.
\end{abstract}


%\tableofcontents

\end{frontmatter}

% \linenumbers


%****************************************************************************************************************
\section{Introduction}
%****************************************************************************************************************

The habitats of animal populations are often geographically divided into many small patches, 
either because of human interference or because natural habitats are patchy. Understanding the dynamics of such populations is a problem of great theoretical and practical interest. Isolated small patches are often extinction prone, for example because of inbreeding in combination with demographic and environmental stochasticity \citep{hanski99}. When the patches are connected into a network by migration, local populations may still be prone to extinction, but the whole population may persist because empty patches are re-colonised 
by migrants from surrounding occupied patches. Moreover, when the migration rate is sufficiently large, local populations can be stabilised by an inflow of immigrants (the rescue effect). Given a model of such a population, the main concern is usually to find out how the different parameter values in the model affect the growth and persistence of the metapopulation.

In Levins' model, the metapopulation dynamics is simplified by treating each patch as 
either occupied or empty \citep{Levins1969}. The rates of patch colonisation and extinction are expressed as functions of the total fraction of occupied patches in the population. In its simplest form
Levins' model describes the time change of the fraction $Q$ of occupied patches:
\begin{equation}
\label{eq:LevinsModel}
\frac{{\rm d}Q}{{\rm d}t} = c\, Q(1-Q) -e\,Q\,.
\end{equation}
Here $c$ is the rate of successful colonisation of an empty patch, and $e$ is the rate at which occupied
patches turn empty (the rate of patch extinction). How these rates are related to the 
life-history parameters determining the stochastic, individual-based
metapopulation dynamics is not explicitly known. In spatially explicit models
the parameters are chosen to be functions of for example the number of occupied neighbouring patches \citep{hui_li04,RoyHarding:2008}.
Eq.~(\ref{eq:LevinsModel}) is generally motivated by assuming a separation of time scales between colonisation and extinction of patches on
the one hand, and the population dynamics within patches on the other hand \citep{Levins1969,hanski_gyllenberg93,Lande1998,Etienne:2002}. 
Many of the mechanisms that can be observed in natural populations (e.g. Allee and rescue effects) 
can be represented by suitable modifications of Eq.~\eqref{eq:LevinsModel}, 
and the qualitative behaviour of such models is well understood 
\citep{Hanski:2000,Etienne:2000,harding_mcnamara02,zhou_wang04}. See also \citet{Zhou:2004,Taylor:2005}, 
and \citet{Martcheva:2007}.

Models that describe the metapopulation dynamics in terms of the fraction of occupied patches
have no explicit connection to the population dynamics within each patch. In other words, 
it is not apparent how individual births, deaths, and migration events translate into colonisation and extinction rates \citep{Etienne:2002}. 

An alternative line of analysis focuses on the population dynamics within a single patch \citep{Drechsler:1997,Lopez:2001,newman_extinction_2004}. The advantage of this approach is that birth and death processes, emigration, and immigration can be modelled explicitly in terms of the number of individuals in the patch. This makes the model much more immediate in terms of biological processes (such as density dependence and Allee effects), but the rest of the metapopulation is reduced to a background source of immigrants (assumed to be stationary) and is not explicitly modelled.

Several authors have attempted to bridge the gap between detailed local population dynamics and the dynamics at the overall population level. \citet{Keeling:2002} estimated 
rates of colonisation and extinction events in Levins' equation from individual-based simulations. 
This method rests on the assumption that Eq.~(\ref{eq:LevinsModel}) is an appropriate description 
of the metapopulation dynamics.

\citet{higgins_metapopulation_2009} 
has investigated the extinction risk of metapopulations 
subject to strong dispersal,
that is, in a limit where migration is faster than the local population dynamics within patches. As pointed out above, the converse
is commonly assumed in writing Eq.~\eqref{eq:LevinsModel}. 
Within his model, \citet{higgins_metapopulation_2009} addressed
the question of how the degree of fragmentation of the metapopulation affects its risk of extinction. 

\citet{Chesson:1981,Chesson1984}, \citet{hanski_gyllenberg93}, \citet{Casagrandi1999,Casagrandi2002}, \citet{Nachman:2000},
\citet{metz_how_2001} and others have analysed the equilibrium states and possible persistence of metapopulations in models 
consisting
of infinitely many patches with local dynamics coupled via a migration pool \citep[reviewed in][]{massol_metapopulation_2009}.
The assumption that there are infinitely many patches is crucial in these studies, as it makes it
possible to pose the question under which circumstances the metapopulation persists {\em ad infinitum}, 
that is, for which choice of parameters 
the metapopulation dynamics reaches a non-trivial stable equilibrium.  
For finite metapopulations, by contrast, there is no stable equilibrium corresponding to persistence 
\citep[for a review of stochastic extinctions in biological populations, see e.g.][]{Ovaskainen2010}.
As is well known, finite populations must eventually become extinct (unless they continue to grow). This fact is referred
to as the \lq merciless dichotomy of population dynamics' by \citet{Jagers}.
Moreover, it remains unclear how the equations employed in these studies are related to Eq.~(\ref{eq:LevinsModel}).

In this paper we characterise the dynamics of metapopulations with a large, but finite, number of patches.  
We derive, from first principles, how the stochastic metapopulation dynamics determines
the distribution of individuals over patches. 
The only critical assumption in our derivation is that the number of patches is large enough.
Especially, it is not necessary to assume that the typical number of individuals per patch is large. Neither is it necessary to assume a time-scale separation between local and migration dynamics. We show that our results
represent the typical transient of the metapopulation towards a quasi-steady state  (if such a state exists). 

On the brink of extinction, that is, close to the bifurcation point where an infinite metapopulation ceases to persist,
we use a systematic expansion of an exact master equation in powers of $N^{-1}$ (where
$N$ is the number of patches) to find the most likely path to extinction, as well as the leading contribution to 
the time to extinction.  
In this case (close to the bifurcation) the metapopulation dynamics simplifies considerably: it can be approximated
by a simple, one-dimensional dynamics. This fact is a consequence of a general principle \citep{GuH83} stating that a dynamical system close
to a bifurcation exhibits a `slow mode': a particular linear combination of the dynamical variables is found to
relax slowly, and the remaining degrees of freedom relax much more quickly and may be assumed to be in local equilibria.
In other words, when the dynamical system is in a perturbed state, the slow mode evolves towards the equilibrium state on a 
longer time-scale than the fast variables do.
This renders the dynamics effectively one-dimensional. We find  the slow mode  of the metapopulation dynamics and show
how it depends on the properties of the local dynamics (given by the local growth rate and the local carrying capacity).
The slow mode determines the stochastic dynamics of finite metapopulations as well as the deterministic dynamics
of metapopulations consisting of infinitely many patches. In the latter case we find that the slow mode obeys 
Eq.~(\ref{eq:LevinsModel}). We derive how the parameters $c$ and $e$ depend on the parameters
determining the life history of the local populations. 
We show, however, that the variable $Q$ is in general not given by the fraction of occupied patches as envisaged by Levins.
But it turns out that $Q$ approaches the fraction of occupied patches in the limit of large carrying capacities 
and large local growth rates (this is the limit of time-scale separation mentioned above).

In other words, we have derived Levins' model, Eq.~(\ref{eq:LevinsModel}),
from a stochastic, individual-based model of a finite metapopulation. We find that Eq.~(\ref{eq:LevinsModel})
is still valid (close to the bifurcation) even when there is no time-scale separation between
the local and the migration dynamics. This is the consequence of the existence of a slow mode.

\citet{Lande1998} have suggested an elegant stochastic generalisation of Le\-vins' model, in an attempt to compute
how the local patch dynamics affects the properties of the quasi-steady state of the metapopulation
(and the average time to extinction of this population). Their 
main idea is to connect the extinction and colonisation rates in Eq.~(\ref{eq:LevinsModel}) to local processes. 
The extinction rate is calculated as the inverse expected time to extinction of 
a single patch at the carrying capacity (which is determined self-consistently), and colonisation is defined as the rate of a single migrant arriving to an empty patch, seeding a population 
that grows to the carrying capacity. 
This scheme is persuasive but not rigorous. The predictions of \citet{Lande1998} have, to our knowledge, never
been tested by comparisons to results of simulations of stochastic, individual-based
metapopulation models. Therefore it is important that our approach allows us to compute the rates of extinction and colonisation from first principles. 
In the limit of large patch carrying capacities and
close to the bifurcation we obtain expressions (exact in the limit we consider) that are very similar, 
but not identical, to the relations proposed by \citet{Lande1998}.

In summary, we characterise the stochastic dynamics of metapopulations
on the brink of extinction. Using an expansion of the exact master equation
describing the stochastic dynamics with the number $N$ of patches as a large parameter, we identify a slow mode in the
metapopulation dynamics, regardless of whether there is a time-scale
separation between local and global dynamics or not. We show under which circumstances
widely used  metapopulation models provide accurate descriptions of metapopulation dynamics.

The remainder of this paper is organised as follows. In section \ref{sec:methods}
we describe the individual-based stochastic metapopulation model investigated here.
We summarise how our numerical experiments were performed and briefly describe
how to represent the metapopulation dynamics in terms of a master equation,
and how to expand this equation in powers of $N^{-1}$.
Our results are described in section \ref{sec:results}. We first discuss the limit $N\rightarrow \infty$,
demonstrate under which circumstances metapopulations persist
in this limit, and analyse the metapopulation dynamics.
Second, we turn to stochastic fluctuations of finite metapopulations, and summarise our
results on fluctuations in the quasi-steady state, the most likely path to extinction as well as the average time to extinction. 
Section \ref{sec:c} contains our conclusions. Appendices A, B, and C summarise details of our calculations.

%****************************************************************************************************************
\section{Methods}
\label{sec:methods}
%****************************************************************************************************************
In this section we define the stochastic, individual-based metapopulation model
that is analysed in this paper. We describe how our numerical experiments are performed
and derive a master equation, Eq.~(\ref{eq:mastereqconstr}), that is the starting point for our mathematical analysis of 
metapopulation dynamics. In general it is not possible to solve this equation
in closed form. We demonstrate how an approximate solution can be obtained
by expanding the master equation using the number $N$ of patches as a large parameter.
In the limit of $N\rightarrow \infty$, the dynamics reduces to a deterministic
model. The equilibrium properties of this model were analysed by 
\citet{Casagrandi1999,Casagrandi2002}, and by \citet{Nachman:2000}. 

%================================================================================================================
\subsection{Stochastic, individual-based metapopulation model}
\label{sec:model}
\begin{figure}[htp]
   \centering
   { \includegraphics[width=0.8\columnwidth]{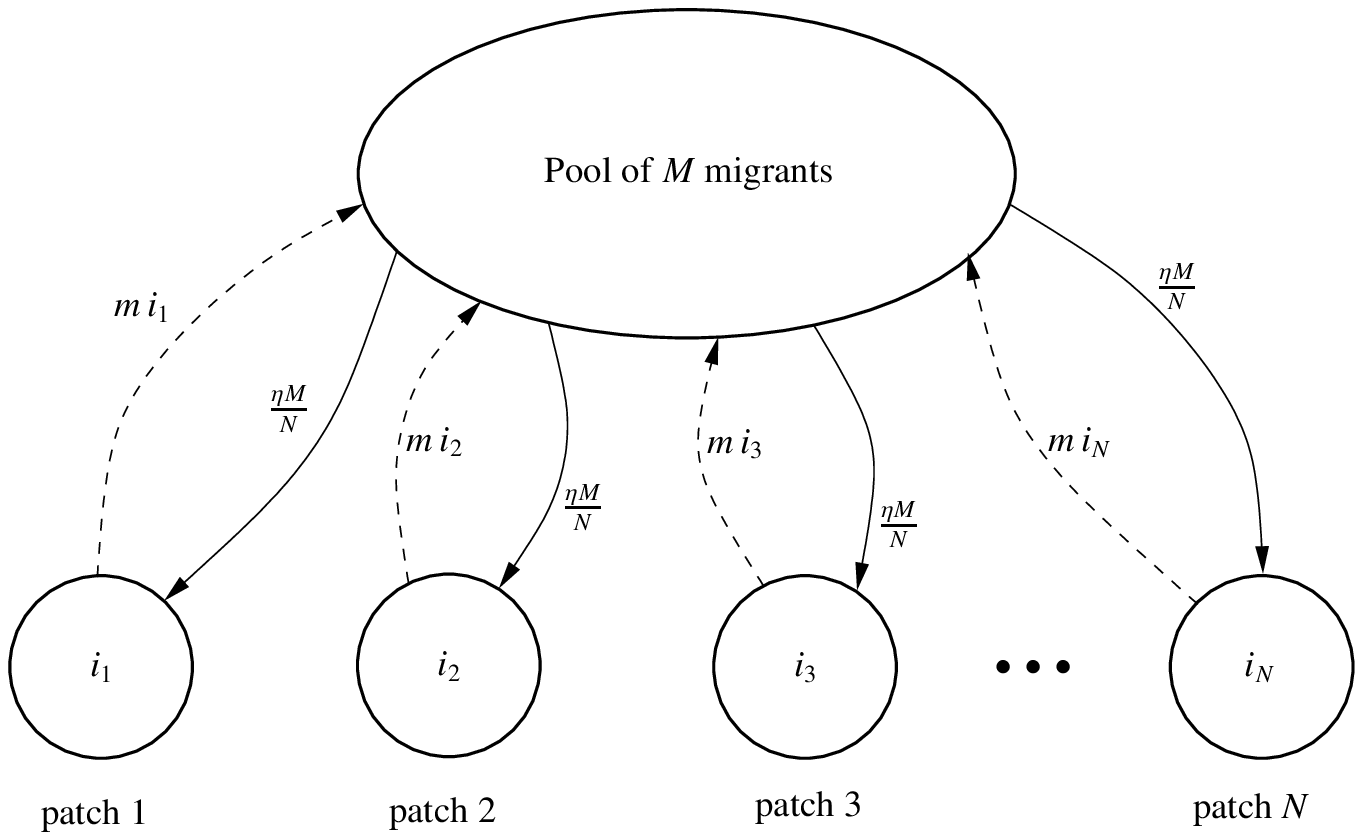}}
      \caption{Illustrates the stochastic, individual-based 
     metapopulation model. The model describes $N$ local populations
     (also referred to as patches). 
     The number of individuals in patch $k$ is denoted by $i_k$. 
     Individuals are born and die with per-capita rates $b_{i_k}$ and $ d_{i_k}$.
     Furthermore, individuals may emigrate to a common
     dispersal pool (emigration rate $m$) where they stay an 
     exponentially distributed time
     (with rate $\eta$) before leaving the pool for one of the $N$ patches.
     Migration may fail if the individuals die before reaching the new patch,
     this possibility is modelled by introducing a death rate $\zeta$ during
     dispersal. The instantaneous number of migrants 
     in the pool is denoted by $M$. The immigration rate from the dispersal pool     
   into any given patch is given by $I=\eta M/N$.}
      \label{fig:1}
\end{figure}

%================================================================================================================
The model consists of a population distributed amongst $N$ 
patches, as illustrated in Fig. \ref{fig:1}. In each patch, the 
local population dynamics is a birth-death process with birth rates $b_i$, and death rates $d_i$. 
Here $i$ denotes the population size in a given patch (and $b_0=d_0=0$).
To simplify the discussion we assume that the rates are the same for all patches,
but more general cases can be treated within the approach described in this paper. In the following
we illustrate our results for a particular choice of
birth and death rates:
\begin{alignat}{2}\label{eq:rates}
 b_i = {}& r\,i           && \text{birth rate}\,,\\
 d_i = {}& \mu i+(r-\mu)i^2/K &\quad& \text{death rate}\,.
\label{eq:dr}
\end{alignat} 
The parameter $r$ is the birth rate per individual, $\mu$ is the density-independent per capita mortality,
and $K$ determines the carrying capacity of a single patch. For simplicity, we take $\mu = 1$ hereafter (this corresponds to measuring time 
in units of the expected life-time of individuals in the absence of density dependence). The parameters occurring
in Eqs.~(\ref{eq:rates},\ref{eq:dr}) are listed in Table~\ref{tab:1}, which summarises the notation used in this article. 

In addition to the local population dynamics, the number of individuals in each patch can change 
because some individuals emigrate from their patch to other patches, or because immigrants arrive from other patches. 

Following 
\citet{hanski_gyllenberg93},  the migration process is modelled as follows: individuals emigrate from a patch at rate $m_i$, 
where $i$ is the population size in the patch in question ($m_0=0$). If the individuals migrate independently and 
with constant rates, $m_i$ is proportional to $i$: 
\begin{alignat}{2}\label{eq:rates2}
  m_i = {}& m\,i           &\quad\quad\quad\quad&\text{emigration rate}\,.
\end{alignat}
More complex migration patterns can be incorporated. If for example individuals moved to avoid overcrowding, 
the emigration rate would be density dependent.
 
The emigrants enter a common dispersal pool, containing the migrants from all patches that have not yet reached their
 target patch. Each migrant stays an exponentially distributed time in the pool, with expected value $1/\eta$, 
before reaching the target habitat, which is chosen with equal probability among all patches. This process is illustrated 
in Fig.~\ref{fig:1}.  Migration may fail if the individuals die before reaching the new habitat; this is modelled by the rate $\zeta$
of dying during dispersal. In practice, the probability of successful migration depends on the background mortality 
of the individuals, on the time the migrating individual spends in the dispersal pool, and on additional perils individuals 
may be exposed to during dispersal (e.g. increased risk of predation due to lack of cover, etc.).
In summary, if there are $M$ migrants, $(\eta + \zeta) M$ individuals leave the dispersal pool per unit of time, and the rate 
of immigration to a given patch is $I = \eta M/N$.

%================================================================================================================
\subsection{Numerical experiments}
\label{sec:nexp}
%================================================================================================================
In the direct numerical simulations, the population evolves in the following manner. First, at any given 
time, the local rates, Eqs.~(\ref{eq:rates}-\ref{eq:rates2}),
sum to a rate of the next locally generated event, $\Lambda_k$, for patch $k$. 
This rate is simply the sum of the local rates, Eqs.~(\ref{eq:rates}-\ref{eq:rates2}).
If patch $k$ contains $i$ individuals then we have:
\begin{equation}
	\Lambda_k = b_i+d_i+m_i\, .
\end{equation}
The sum of these rates over all patches, $\Lambda = \sum_k \Lambda_k$, for $k=1,\ldots,N$,  
yields the rate for the next event occurring in the population.
Thus the simulation proceeds by generating an exponentially distributed random number with expected value $1/\Lambda$. Second, at each time step, 
a patch is chosen with probability $\Lambda_k/\Lambda$. Third, the type of event is chosen randomly:
birth with probability  $b_i/\Lambda_k$, death with probability $d_i/\Lambda_k$ or emigration with $m_i/\Lambda_k$.
Fourth, the numerical experiments described below were performed in the limit $\eta\rightarrow \infty$ and for $\zeta=0$.
This means that emigrating individuals are immediately assigned to a randomly chosen patch (possibly the one they come from).  

%================================================================================================================
\subsection{Master equation}\label{sec:mastereq}
%================================================================================================================
A natural and commonly adopted approach to describe stochastic population dynamics is to derive 
a master equation \citep{vKa81} for the change in time of the probability $\rho$ of observing
$i_1$ individuals in the first patch, $i_2$ individuals in the second patch, $\ldots,$ and of observing $M$ individuals in the migrant pool.
Recently this approach was adopted by \citet{Meerson2011} to describe 
the dynamics of local birth-death processes coupled by nearest-neighbour interactions (diffusion).

In the following we pursue a different approach. 
Since the local population dynamics is assumed to be the same within all patches, 
it is sufficient to count the number of patches with a given number of individuals rather than keeping track of the number of individuals in each patch.  Let $n_j$ denote the number of patches with $j$ inhabitants at a given time. 
The state of the population is described by the variables $n_0,n_1,\ldots,$ and $M$.
In the master equation, the time derivative of the probability $\rho(n_0, n_1, \ldots,M;t)$
of observing the system in a given state $n_0, n_1, \ldots,M$ at time $t$,
is the rate of arriving to a given state from other states, minus the rate of transitions to other states.
Thus, we find the master equation for the probability $\rho(n_0,n_1,\ldots,M;t)$ by considering all possible transitions
between the states of the population, contributing to the change of $\rho(n_0,n_1,\ldots,M;t)$.

For example, consider the effect of local births on the probability of finding the system in a given, 
\lq focal' state $n_0, n_1, \ldots,M$ at time $t$. A birth event in a patch with $j$ individuals corresponds to the transition $n_j \rightarrow n_j - 1$ and $n_{j+1} \rightarrow n_{j+1} + 1$. Thus, $\dd\rho/\dd t$ has the contribution
\begin{equation}
-b_j n_j \rho(n_0, n_1, \ldots,M;t)
\end{equation}
from births in the focal state. This contribution 
is negative since any such birth in the focal state moves the system away from this state.
In order to obtain the positive contributions to  $\dd\rho/\dd t$, we calculate the rate of 
arriving to the focal state by a birth in a different state:
\begin{equation}
	b_j(n_{j} + 1)\rho(\ldots, n_j + 1, n_{j+1} - 1, \ldots) 
\end{equation}
A compact way of writing these contributions is in terms of the raising and lowering operators $\mathbb{E}_{j}^\pm$, defined by their effects 
on a function $g$ \citep{vKa81}:
\begin{align}
\label{eq:rl}
        \mathbb{E}_{j}^+ g(\dotsc,n_j,\dotsc) ={}& 
     g(\dotsc, n_j + 1, \dotsc)\,,\\  
     \mathbb{E}_{j}^- g(\dotsc,n_j,\dotsc) ={}& g(\dotsc, n_j - 1, \dotsc) \nn \,.
\end{align}
In our example, the total contribution to $\dd\rho/\dd t$ from birth events can thus be written as
\begin{equation}
	\sum_{j=0}^\infty (\mathbb{E}_j^+\mathbb{E}_{j+1}^- - 1) b_{j} n_j \rho \,.
\end{equation}
Adding up the contributions due to death, emigration, and immigration we find:
\begin{align}
\frac{\dd \rho}{\dd t}
={} & \sum_{j=0}^\infty (\mathbb{E}_j^+\mathbb{E}_{j+1}^--1) b_{j} n_j \rho\nonumber\\
& + \sum_{j=0}^\infty(\mathbb{E}_j^+\mathbb{E}_{j-1}^
--1) d_{j} n_j \rho  + \sum_{j=0}^{\infty}\left(\mathbb{E}_{j}^{+}\mathbb{E}_{j-1}^{-}\mathbb{E}_{M}^{-}-1\right)m_{j}n_{j}\rho \nn \\ \label{eq:mastereq}
&+\sum_{j=0}^{\infty
}\left(\mathbb{E}_{j}^{+}\mathbb{E}_{j+1}^{-}\mathbb{E}_{M}^{+}-1\right)\eta M\frac{n_{j}}{N}\rho + \left(\mathbb{E}_{M}^{+}- 1\right)\zeta M\rho\, .
\end{align}
To simplify the discussion we assume that migration is instantaneous,
this corresponds to taking the limit $\eta\rightarrow \infty$ and $\zeta=0$.
In this limit, the immigration rate to a patch is given by:
\begin{equation}
\label{eq:I1}
        I = \frac{1}{N}\sum_{j=0}^{\infty}m_j n_j\,,
\end{equation} 
and the corresponding master equation takes the form:
\begin{align}
\frac{\dd \rho}{\dd t}
\nn
= {} & \sum_{j=0}^\infty (\mathbb{E}_j^+\mathbb{E}_{j+1}^--1) b_{j} n_j \rho + \sum_{j=0}^\infty(\mathbb{E}_j^+\mathbb{E}_{j-1}^--1) d_{j} n_j \rho\\
&+ \frac{1}{N}\sum_{i=1}^\infty \sum_{j=0}^\infty
\mathbb{E}_{i-1}^- \mathbb{E}_i^+\mathbb{E}_j^+\mathbb{E}_{j+1}^- m_i n_i(n_j - \delta_{ij} + \delta_{i-1j}) \rho
- \sum_{i=1}^\infty m_i n_i \rho \, .
\label{eq:mastereq2}
\end{align}
The last two terms on the right-hand side of Eq.~(\ref{eq:mastereq2})  describe instantaneous migration where $M=0$.
The terms involving Kronecker $\delta$-symbols (see Table~\ref{tab:1}) on the right-hand side of Eq.~\eqref{eq:mastereq2} arise from enforcing that, for the migration of an individual, emigration must precede immigration.  These terms are of higher order in $N^{-1}$. 

The number of patches is given by $N = \sum_{j=0}^\infty n_j$.
The following discussion is simplified by making this constraint explicit 
in the master equation (\ref{eq:mastereq2}). This is achieved by considering
$n_0$, the number of empty patches, to be be a function of $n_1,n_2,\ldots$:
\begin{equation}
\label{eq:c3}
n_0 = N-\sum_{j=1}^\infty n_j\,.
\end{equation}
Using Eq.~(\ref{eq:c3}) we find the following master equation:
\begin{align} \label{eq:mastereqconstr}
	\frac{\dd \rho(\ve n,t)}{\dd t} &{}= 
	\sum_{j=1}^\infty (\mathbb{E}_j^+\mathbb{E}_{j+1}^--1) b_{j} n_j \rho(\ve n,t) 
	+ \sum_{j=1}^\infty(\mathbb{E}_j^+\mathbb{E}_{j-1}^--1) d_{j} n_j \rho(\ve n,t) \nn\\
	+& \frac{1}{N}\sum_{i=1}^\infty \sum_{j=1}^\infty  (\mathbb{E}_{i-1}^- \mathbb{E}_i^+ \mathbb{E}_j^+\mathbb{E}_{j+1}^- - 1) m_i n_i (n_j - \delta_{ij} + \delta_{i-1j}) \rho(\ve n,t) \nn \\
		+& \sum_{i=1}^\infty (\mathbb{E}_{i-1}^- \mathbb{E}_i^+ \mathbb{E}_{1}^- - 1) m_i n_i \Big(1-  \frac{1}{N}\sum_{k=1}^\infty n_k  + \delta_{i1}\Big) \rho(\ve n,t) 
		- \sum_{i=1}^\infty m_i n_i \rho \, .
\end{align}
Here the components $n_j$ of the vector $\ve n = (n_1,n_2,\ldots)^{\sf T}$ denote
the number of patches with $j\geq 1$ individuals, and $\mathbb{E}_{0}^\pm\equiv 1$. 
Eq.~(\ref{eq:mastereqconstr}) describes the stochastic population dynamics of the metapopulation model
considered here.  

In the next section we describe the approximate method of solving Eq.~(\ref{eq:mastereqconstr}) adopted in the following. 
It corresponds to a systematic expansion of the master equation,
using the number $N$ of patches in the population as a large parameter.
This approach does not require that the number of individuals per
patch is large, neither is it necessary to assume a time-scale separation
between local population dynamics and migration. Our approach makes
it possible to accurately describe the dynamics of metapopulations
consisting of many small patches. In the limit of infinitely many patches,
to lowest order in the expansion, a deterministic metapopulation dynamics is obtained. 
Stochastic fluctuations in metapopulations with a large
but finite number of patches are described by the leading order
of the expansion.

%================================================================================================================
\subsection{Expansion of the master equation}
\label{sec:exp}
%================================================================================================================

When $N$ is large we expect that the probability $\rho$ of observing a given distribution of $\ve n$ 
changes only little when patches change in population size. It is important to emphasise that no assumption is made
concerning the size of the changes to the population size of any single habitat. 
It is merely assumed that there are sufficiently many patches  that they form a statistical ensemble.
In this case it is perfectly possible to have big jumps in the population sizes of individual patches in the model 
without violating the assumption that the distribution of $\ve n$ is smooth,
and to allow the local population dynamics to depend sensitively on the local patch population size
when there are few individuals in the patch ---  modelling for example 
the effect of abundance on mating success \citep{SaetherEngen:2004, MelbourneHastings:2008}.

It is convenient to express the the components of the vector
 $\ve n$ in terms of the scaled frequencies $f_j = n_j/N$.
The probability $\widetilde\rho$ of observing $f_1, f_2, \ldots$ 
is related to the probability $\rho$ by
\begin{equation}
\label{eq:N1}
        \widetilde{\rho}(\ve f;t) = \rho(N\ve f;t)\, .
\end{equation}
In contrast to the distribution of $n_j$, 
the distribution of $f_j$ is expected to become approximately independent of $N$ for a large number of patches.  
A change of a single count $n_j$ leads only to a change of $1/N$ in $f_j$. 
Since the probability $\widetilde\rho$ is an approximately continuous function of $f_j$ in the limit of large
values of $N$, we seek an approximate solution of Eq.~(\ref{eq:mastereqconstr}) by expanding this
equation in powers of $N^{-1}$.

In the limit of $N\rightarrow \infty$ we expect that stochastic fluctuations are negligible,
so that the metapopulation dynamics becomes deterministic. It takes the form of a kinetic equation for $\ve f$:
\begin{equation}
\label{eq:det}
\frac{{\rm d}\ve f}{{\rm d}t} = \ve v(\ve f)\,.
\end{equation}
The right-hand side of this equation, $\ve v(\ve f)$, depends upon the details of the model analysed.
For the model described in Section~\ref{sec:model}, the form of $\ve v(\ve f)$ is given in Eq.~(\ref{eq:v}).
In the limit of $N\rightarrow\infty$, the question whether or not the metapopulation may persist is answered by
finding the steady states $\ve f^\ast$ of the system (\ref{eq:det}), given by $\ve v(\ve f^\ast)= \ven 0$.
Persistence corresponds to the existence of a stable steady state with positive components $f_j^\ast >0$
for some values of $j$. If, by contrast, the only stable steady state is $\ve f^\ast =\ven 0$, then
the metapopulation will definitely become extinct. The time to extinction is determined
by Eq.~(\ref{eq:det}), and by the initial conditions. 

For large (but finite) values of $N$ the metapopulation fluctuates around the stable steady
states $\ve f^\ast$ of Eq.~(\ref{eq:det}). These fluctuations can be described by expanding
the master equation (\ref{eq:mastereqconstr}) to leading order in $N^{-1}$. The stochastic fluctuations
are expected to be small when $N$ is large, but they are crucial to the metapopulation dynamics as mentioned in the Introduction.
When the number of patches is finite, the metapopulation must eventually become extinct
(the state $\ve f= \ven 0$ is the only absorbing state of the master equation). When $N$ is large,
the time to extinction from a stable steady state of the deterministic dynamics is
expected to be large. By analogy with standard large-deviation analysis, we expect
that the time to extinction increases exponentially with increasing $N$, giving
rise to a long-lived quasi-steady state. Below we analyse its properties, and 
estimate the time to extinction of the metapopulation. Our results are consistent with the above expectation,
and we show that the time to extinction depends sensitively on the parameters of
the model ($r, K, m$, and $N$).

%----------------------------------------------------------------------------------------------------------------
\subsubsection{Deterministic dynamics in the limit of $N\rightarrow\infty$}\label{subs:detdyn}                   
%----------------------------------------------------------------------------------------------------------------
To simplify the notation, we drop the tilde in Eq.~(\ref{eq:N1}) so that $\rho(\ve f,t)$ is the probability 
of observing,
at time $t$,
a fraction $f_1$ of patches with one individual, a fraction $f_2$ of patches with two individuals, and so forth.
In deriving the expansion of the master equation (\ref{eq:mastereqconstr}) we use the approach 
described by \citet{vKa81}. 
Assuming that $\rho$ is a smooth function of $\ve f$, the action of the raising and lowering operators,
Eq.~(\ref{eq:rl}), can be written as
\begin{equation}
\label{eq:vK}
\mathbb{E}_j^\pm = \exp(\pm N^{-1} \partial_{f_j})\,.
\end{equation}
The master equation (\ref{eq:mastereqconstr}) is expanded as follows.
We replace $n_j$ by $N f_j$ in Eq.~(\ref{eq:mastereqconstr}), insert Eq.~(\ref{eq:vK}), and 
expand in powers of $N^{-1}$. Keeping only the lowest order in $N^{-1}$,
we arrive at an equation for $\rho$ that corresponds to 
deterministic dynamics of the form (\ref{eq:det}). We find that the components of $\ve v(\ve f)$
are given by:
\begin{alignat}{2}
v_j(\ve f)= {}& (b_{j-1}+I) f_{j-1} + (d_{j+1}+m_{j+1}) f_{j+1}  &&\label{eq:v} \\
&- (b_j+I+d_j+m_j)f_j &&\quad\mbox{for $j>1$}\,,\nn\\
v_1(\ve f) = {}& I(1-\sum_{k=1}^\infty f_k) + (d_{2}+m_{2}) f_{2} -(b_1+I+&&d_1+m_1)f_1 \nn \,.
\end{alignat}
Here
$I=\sum_{k=1}^\infty m_kf_k$
is the rate of immigration into a given patch, corresponding to Eq.~(\ref{eq:I1}).
Since $I$ depends upon $\ve f$, the deterministic dynamics (\ref{eq:det},\ref{eq:v})
is nonlinear. We note that 
Eqs.~(\ref{eq:det},\ref{eq:v}) correspond to the metapopulation
model suggested by \citet{Casagrandi1999} and \citet{Nachman:2000}. Here we have derived it by a systematic expansion
of the exact master equation in powers of $N^{-1}$ where $N$ is the number of patches. 
Our derivation emphasises the fact that Eqs.~(\ref{eq:det},\ref{eq:v}) approximate the metapopulation dynamics by 
a deterministic equation. This approximation improves as $N$ becomes larger, and 
becomes exact as $N \to \infty$. \citet{arrigoni2003das} has shown that this limit 
holds under quite general assumptions for the underlying stochastic model, namely that 
the time-evolution of the probability measure over the states of the model 
converges to the deterministic time evolution as $N\rightarrow\infty$.
\citet{Casagrandi1999}, \citet{Nachman:2000}, and others have studied how the stability of the steady 
states of the deterministic dynamics depends upon the parameters of the model.
However, Eqs.~(\ref{eq:det},\ref{eq:v}) cannot be used to 
determine how stochastic population dynamics affects metapopulation persistence. In order to take the stochastic fluctuations into account, it is necessary to consider the next  order in $1/N$.

%----------------------------------------------------------------------------------------------------------------
\subsubsection{Quasi-steady state distribution at finite but large values of $N$}\label{sec:qssfinite}
%----------------------------------------------------------------------------------------------------------------
\label{sec:qss}
\begin{figure}[htp]
\centering
   \includegraphics[width=0.4\columnwidth,height=0.4\columnwidth]{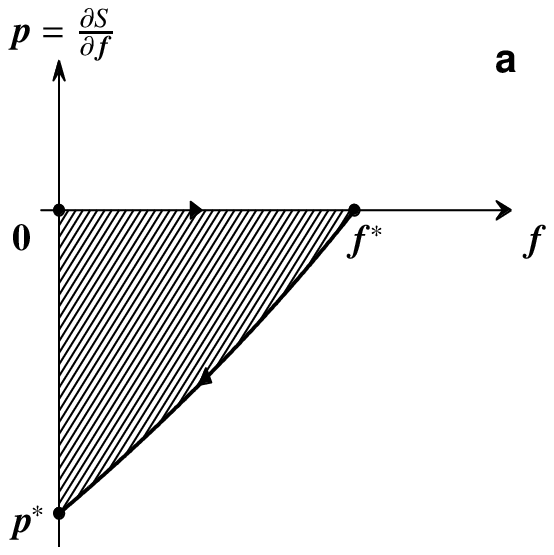}\\[0.5cm]
   \hspace{-2pt}\includegraphics[width=0.4\columnwidth,height=0.4\columnwidth]{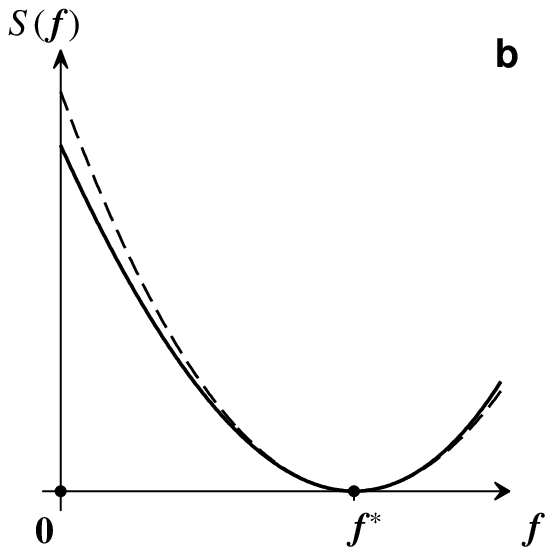}   
\caption{\label{fig:fp_plane} Illustrates the WKB method. 
 (\textsf{\textbf{a}}) Schematic plot of the $\ve f$-$\ve p$ plane.
 The significance of the variables $\ve f$ and $\ve p$ is explained in Section \ref{sec:exp}
 The deterministic dynamics, Eq.~(\ref{eq:det}), corresponds to motion along the $\ve f$-axis.
 The form of the action $S(\ve f)$, Eq.~(\ref{eq:action}), is determined by the path from the quasi-steady state $(\ve f^*, \ven 0)$ 
 to the fluctuational extinction point $(\ven 0, \ve p^*)$, corresponding
to the most likely path to extinction. This is a standard
 situation, often referred to in the literature. See for example Fig.~1 in Box~3 in \citep{Ovaskainen2010}.
 (\textsf{\textbf{b}}) Schematic. Shows  $S(\ve f)$ as a function of $\ve f$ (solid line).
 Also shown (dashed line) is a quadratic approximation of $S(\ve f)$. This illustrates
 that the quasi-steady state distribution $\rho(\ve f)$ is Gaussian close to $\ve f^*$, but in general non-Gaussian in the tails.}
\end{figure}
Consider a stable steady state of the metapopulation in the limit of
infinitely many patches, that is, a stable steady state $\ve f^\ast$ of the deterministic
dynamics (\ref{eq:det}). 

Metapopulations consisting of a finite number $N$ of patches exhibit random 
fluctuations caused by the random sequence of birth-, death-, and migration
events. When the number of patches is large, these fluctuations
are expected to be small, and the components $f_j$ of the vector $\ve f$ are expected
to fluctuate closely around those of the stable steady state $\ve f^\ast$.
If the fluctuations around $\ve f^\ast$ are small, the
metapopulation may persist for a very long time (but extinction of 
this finite metapopulation is certain, as explained above.).
This situation
is commonly referred to as a \lq quasi-stable' steady state. 
We analyse its properties by expanding the master equation to leading
order in $N^{-1}$, using a standard method that is referred to as \lq WKB analysis' \citep{Wilkinson07},
as the \lq eikonal approximation' \citep{Dyk94}, or as the \lq large-deviation principle' 
in the mathematical literature \citep{Frei84}.

We briefly outline this method in the remainder of this subsection. For a comprehensive
description, the reader is referred to \citet{Altland10}, \citet{Elgart04}, and \citet{Dyk94}.
The method has been successfully used to describe fluctuations in finite biological populations,
describing for example the spreading of epidemics \citep{Dyk08} or the risk of extinction
of biological populations (see \citet{Ovaskainen2010} for a review). 

In the quasi-steady state we expect ${\rm d}\rho/{\rm d}t \approx 0$ and seek a solution of the master equation of the form
\begin{equation}
\label{eq:ansatz}
\rho(\ve f) \approx \exp[- NS(\ve f) + \mbox{higher orders in $N^{-1}$}]\,.
\end{equation}
The function $S(\ve f)$ is commonly referred to as the \lq action'. 
It not only depends upon $\ve f$, but also on the parameters of the problem ($r$, $K$, and $m$ in our case).
This dependence is not made explicit in Eq.~(\ref{eq:ansatz}).
In the limit of a large number of patches, this function determines the form 
of the probability distribution $\rho(\ve f)$ in the quasi-steady state, and its sensitive dependence upon
the parameters $r$, $K$, and $m$.

As mentioned above, $\rho(\ve f)$ is expected to be strongly peaked at $\ve f^\ast$
in the limit of large values of $N$. In other words,
the quasi-steady state distribution concentrates on the deterministic fixed point $\ve f^\ast$ 
as the number of patches tends to infinity. It is therefore convenient to define
the action function such that $S(\ve f^\ast)=0$. When the number of patches is large,
we expect the distribution $\rho(\ve f)$ to be Gaussian
in the vicinity of the steady state. Correspondingly,  $S(\ve f)$ 
is expected to be approximated by a quadratic function of $\delta\ve f=\ve f - \ve f^\ast$. 
By contrast, non-Gaussian tails of this distribution, corresponding to large deviations of $\ve f$ from $\ve f^\ast$,
reflect the particular properties of the extinction dynamics of the metapopulation.

The form of the function $S(\ve f)$ is determined by 
inserting the ansatz (\ref{eq:ansatz}) into the master equation (\ref{eq:mastereqconstr}), making
use of Eq.~(\ref{eq:vK}), and expanding 
in $N^{-1}$. One finds a first-order partial differential equation for $S(\ve f)$:
\begin{equation}
\label{eq:HJE}
	 0  =  H\left(\ve f,\frac{\partial S(\ve f)}{\partial \ve f}\right).
\end{equation}
In our case, for the master equation (\ref{eq:mastereqconstr}), we obtain:
\begin{align}
\nn H(\ve f, \ve p) ={}& 
\sum_{j=1}^\infty ({\rm e}^{p_{j+1}-p_j}-1) b_j f_j + \sum_{j=1}^\infty({\rm e}^{p_{j-1}-p_j}-1)d_j f_j\\
&+\sum_{i=1}^\infty \sum_{j=1}^\infty ({\rm e}^{p_{i-1}-p_i -p_j + p_{j+1}}-1) m_i f_i f_j \label{eq:H}\\
&+\sum_{j=1}^\infty ({\rm e}^{p_{j-1}-p_j + p_{1}}-1) m_j f_j \Big(1-\sum_{k=1}^\infty f_k\Big)\,.
\nn          
\end{align}
Here $\ve p = (p_1,p_2,\ldots)^{\sf T}$, $p_j = \partial S/\partial  f_j$ for $j \geq 1$, and $p_0 = 0$. 

The  solution of Eqs.~(\ref{eq:HJE},\ref{eq:H}) is found by recognising
that Eq.~(\ref{eq:HJE}) is a so-called \lq Hamilton-Jacobi equation' \citep{Frei84}.
As a consequence, the action $S(\ve f)$ can be determined by
solving the set of equations
\begin{equation}
\label{eq:HEM}
\frac{{\rm d}\ve f}{{\rm d}t} = \frac{\partial H}{\partial \ve p}\,,\quad\mbox{and}\quad
\frac{{\rm d}\ve p}{{\rm d}t} = -\frac{\partial H}{\partial \ve f}\,.
\end{equation} 
These equations have the form of Hamilton's equations in
classical mechanics with configuration-space variables $\ve f$.
The variables 
\begin{equation}
\ve p = \frac{\partial S}{\partial \ve f}
\end{equation}
are therefore referred to as \lq momenta'.
An important property of Eq.~(\ref{eq:HEM}) is that
the function $H(\ve f(t), \ve p(t))$ remains constant
along the \lq trajectories' $\big(\ve f(t),\ve p(t)\big)$, that
is, along the solutions of Eq.~(\ref{eq:HEM}).

How is the form of $S(\ve f)$ determined by these solutions?
The quasi-steady state solution of Eq.~(\ref{eq:mastereqconstr})
corresponds to solutions of Eq.~(\ref{eq:HEM}) satisfying
\begin{align}
&\left . 
\begin{array}{l}
\ve f(t) \rightarrow \ve f^\ast \\
\ve p(t) \rightarrow \ven 0
\end{array}
\right\} 
\mbox{as $t\rightarrow -\infty$}\,,\quad \ve f(t) \rightarrow \ve f\quad \mbox{as $t\rightarrow \infty$}\,, \nn\\[0.5em]
&\;\; \mbox{and}\quad\mbox{$H\big(\ve f(t),\ve p(t)\big)=0$}\label{eq:bc}\,.
\end{align}
To every such solution corresponds an action $S(\ve f)$ 
obtained by integrating the momentum along the path $\big(\ve f(t),\ve p(t)\big)$:
\begin{equation}
\label{eq:action}
S(\ve f) = \int_{-\infty}^\infty\!\!\!{\rm d}t\, \ve p^{\sf T} \frac{{\rm d}{\ve f}}{{\rm d}t}\,.
\end{equation}
Here $\ve p^{\sf T}$ denotes the transpose of the vector $\ve p$.

The boundary conditions (\ref{eq:bc}) are motivated as follows. Eq.~(\ref{eq:HJE}) enforces
$H=0$. Further, observe that 
the stable steady state $\ve f^\ast$ of the deterministic dynamics (\ref{eq:det}) corresponds to a 
steady state $(\ve f^\ast,\ven 0)$ of Eq.~(\ref{eq:HEM}).
This can be seen by expanding the function $H(\ve f,\ve p)$ to second order in $\ve p$
around $\ve p = \ven 0$
\begin{equation}
\label{eq:H2}
H(\ve f, \ve p) = \ve p\tr\ve v(\ve f) + \frac{1}{2}\ve p \tr {\ma D}(\ve f) \ve p+\cdots\,.
\end{equation}
Here $\ve v(\ve f)$ is given by Eq.~(\ref{eq:v}), and the symmetric matrix ${\ma D}(\ve f)$ 
has elements $D_{ij} = \partial^2H/\partial p_i\partial p_j$. The elements 
are given in appendix \ref{app:A}. Eq.~(\ref{eq:H2}) shows that
the dynamics for $\ve p=\ven 0$ corresponds to the deterministic dynamics
given in Eq.~(\ref{eq:det}). The stability of the steady state $(\ve f^\ast, \ven 0)$
is determined by the eigenvalues of the matrix
\begin{equation}
\label{eq:J}
{\bf J} = \left(\!\! \begin{array}{cc}{\ma A} & {\ma D} \\{\ma 0} & -{\ma A}^{\sf T}\end{array} \!\! \right)\,.
\end{equation}
The matrix ${\ma A}(\ve f)$ has elements $A_{ij} = \partial v_i/\partial f_j$ evaluated at $\ve f = \ve f^\ast$.
It is the stability matrix of the stable steady state $\ve f^\ast$ of the deterministic dynamics (\ref{eq:det}).
Its elements can be obtained from Eq. (\ref{eq:A}) in appendix \ref{app:A}.  Similarly, ${\ma D}$ is evaluated at $\ve f^\ast$.
Assuming that the steady state $\ve f^\ast$ is stable, 
the eigenvalues $\lambda_\alpha$ of ${\ma A}$ 
must have negative real parts. In our case it turns out that the eigenvalues are in fact negative. We write $0 > \lambda_1 > \lambda_2 > \cdots $. The eigenvalues
of ${\bf J}$ occur in pairs $\lambda_\alpha$, and $-\lambda_\alpha$. The steady state $(\ve f^\ast,\ven 0)$
of Eq.~(\ref{eq:HEM}) is thus a saddle. In other words, stochastic fluctuations allow metapopulations 
consisting of a finite number of patches to escape from the steady state $\ve f^\ast$ (that
is stable in the limit $N\rightarrow\infty$) to extinction ($\ve f = \ven 0$).  
In general there are (infinitely) 
many such paths satisfying the boundary conditions (\ref{eq:bc}).
\citet{Frei84} formulated a variational principle for the most likely escape path: in the limit
of large values of $N$, the metapopulation goes extinct predominantly along this
path. The quasi-steady state distribution reflects this property: configurations $\ve f$
along this path are assumed with higher probability.
According to the principle described by \citet{Frei84}, the most likely 
escape path is the one with extremal action, Eq.~(\ref{eq:action}).

The picture summarised above is schematically depicted in Fig.~\ref{fig:fp_plane},
for the case where the metapopulation persists in the limit of $N\rightarrow \infty$.
Fig.~2a shows the $\ve f$-$\ve p$ plane. 
As explained above, the point $(\ve f^\ast, \ven 0)$
corresponds to the stable steady state of the deterministic dynamics,
where the infinitely large metapopulation persists. The point 
 $(\ven 0, \ven 0)$ corresponds to extinction, it is unstable.
Solving Eq.~(\ref{eq:HEM}) and inserting $\ve p=\ven 0$ yields the deterministic 
dynamics (\ref{eq:det}), connecting these two fixed points. 
In the limit of $N\rightarrow \infty$, the metapopulation dynamics is constrained to the $\ve f$-axis
and must approach $\ve f^\ast$, as the arrow on the $x$-axis
in Fig.~\ref{fig:fp_plane}a indicates. In other words, in the situation depicted
in Fig.~\ref{fig:fp_plane}a extinction never occurs in the limit of $N\rightarrow \infty$.  

In finite metapopulations, the situation is entirely different. Fig.~\ref{fig:fp_plane}a 
illustrates that there is a path from $\ve f^\ast$ to $\ve f=\ven 0$,
reaching $\ve f=\ven 0$ at the so-called \lq fluctuational extinction point'
$(\ven 0,\ve p^\ast)$. Along this path, the momenta $\ve p(t)$ assume non-zero
values. These variables characterise the sensitivity of the finite metapopulation
to stochastic fluctuations ($\ve p=\ven 0$ in the deterministic limit $N\rightarrow\infty$
as mentioned above). The form of $S(\ve f)$ (Fig.~\ref{fig:fp_plane}b) is determined by evaluating
Eq.~(\ref{eq:action}) along this path, satisfying conditions (\ref{eq:bc}).
In the limit of large values of $N$,
the average time to extinction of the metapopulation scales as \citep{Dyk94}
\begin{equation}
\label{eq:Text0}
T_{\rm ext} = A \exp\big[N S(\ve f=\ven 0)\big]\,.
\end{equation}
The coefficient $A$ may depend on $N$, as well as $r, K$,and $m$.
In the argument of the exponent, only the leading $N$-dependence is explicit in Eq. (\ref{eq:Text0}).
The argument of the exponential determines the sensitive dependence
of the time to extinction. In one-dimensional problems with a single component $f$
(the case discussed in the review by \citet{Ovaskainen2010}),
Eq.~(\ref{eq:action}) shows that $S(0)$ is given by the shaded area in Fig.~\ref{fig:fp_plane}a.
In the case of our metapopulation model, by contrast, the
vector $\ve f$ has infinitely many components. It is therefore not possible, in general, 
to find the most likely path 
from $\ve f^\ast$ to the fluctuational extinction point explicitly.
In practice one may truncate the dynamics by only considering a finite number of variables
$\ve f_j$ (up to a maximal value of $j_{\rm max}$). This is expected to be a good
approximation when $j_{\rm max}$ is taken to be much larger than the carrying capacity $K$,
since $f_j \approx 0$ for $j \gg K$. In our subsequent analysis of the problem
we make use of the fact that the dynamics simplifies considerably
in the vicinity of a bifurcation of the deterministic dynamics \citep{GuH83,Dyk94}.

As mentioned above, near the steady state $\ve f^\ast$, the distribution (\ref{eq:ansatz}) is expected
to be Gaussian. This corresponds to an action quadratic 
in $\delta \ve f = \ve f-\ve f^\ast$:
\begin{equation}
\label{eq:S2}
S(\ve f) \approx \frac{1}{2}\delta\ve f^{\sf T} {\ma C}^{-1}  \delta\ve f\,.
\end{equation}
The matrix ${\ma C}$, which is the covariance matrix of the distribution \eqref{eq:ansatz} 
multiplied by $N$, can be obtained from the linearised dynamics, Eq.~(\ref{eq:HEM}).
Using $\ve p = \partial S/\partial \ve f$ we have
\begin{equation}
\label{eq:bc0}
\delta \ve p 
= {\ma C}^{-1} \delta \ve f \,.
\end{equation}
According to Eq.~(\ref{eq:bc}), the dynamics must obey $H\big(\ve f(t),\ve p(t)\big)=0$.
It follows from Eq.~(\ref{eq:H2}) that the linearised dynamics satisfies this constraint provided 
\begin{equation}
\label{eq:C2} {\ma A}{\ma C} + {\ma C}{\ma A}^{\sf T} +{\ma D} = {\ma 0}\,.  
\end{equation} 
This equation determines the covariance matrix ${\ma C}$ in Eq. \eqref{eq:S2} in terms of the matrices ${\ma A}$
and ${\ma D}$.
%****************************************************************************************************************
\section{Results and discussion}
\label{sec:results}
%****************************************************************************************************************
In this section we summarise our results for the population dynamics of the metapopulation model
described in Section \ref{sec:model}, using the expansion of the master equation outlined
in Section \ref{sec:exp}. This section is divided into two parts. We first discuss the 
limit of infinitely many patches. 
Second, we analyse the most likely path to extinction in finite metapopulations.
We also   summarise our results for the average time to extinction of the metapopulation,
and demonstrate that it depends sensitively on upon the parameters of the model
($r, K, m$, and $N$).
\begin{figure}[htp]
\centering
   \includegraphics[width=0.39\columnwidth]{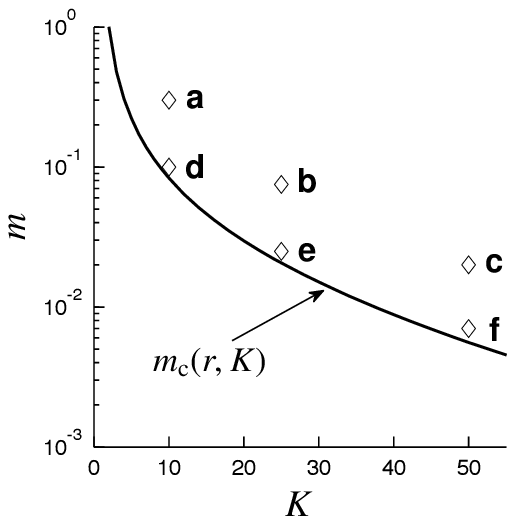}\qquad
   \includegraphics[width=0.54\columnwidth]{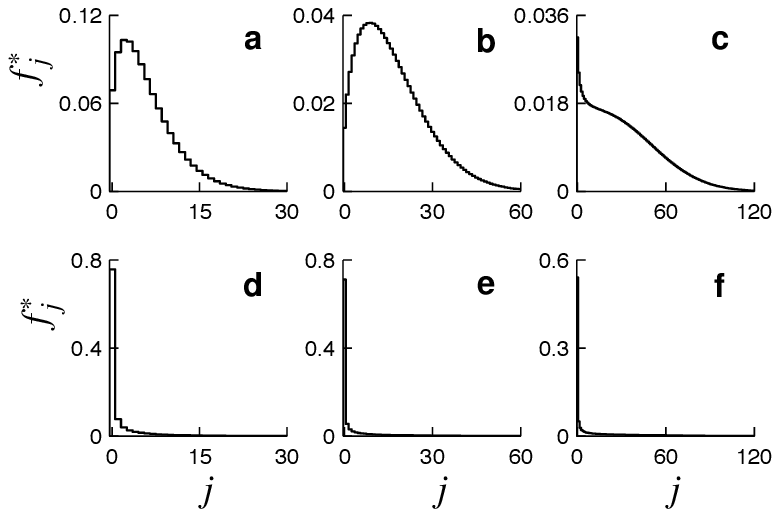}
   \caption{\label{fig:mc} Left:
   critical emigration rate $m_{\rm c}$ in the limit of infinitely many patches, $N\rightarrow \infty$,
as a function of the carrying capacity $K$ (solid line), computed from Eq.~(\ref{eq:mc}).
The growth rate is $r=1.05$. Above this critical line, metapopulations persist in the limit of $N\rightarrow \infty$.    Right: six different stable steady states $\ve f^\ast$, obtained by solving Eqs.~(\ref{eq:fast}, \ref{eq:sc}) numerically, for $r=1.05$, and for the values
   of the emigration rate $m$ and $K$ indicated in the left panel.}
\end{figure}

%================================================================================================================
\subsection{Infinitely many patches}
\label{sec:detdyn}
%================================================================================================================
It was shown in Section \ref{sec:exp} that in the limit of infinitely many patches, the metapopulation
dynamics is described by the deterministic equation (\ref{eq:det}), corresponding
to the model proposed by \citet{Casagrandi1999} and \citet{Nachman:2000}. 
In this subsection we briefly summarise our results on the persistence and
the relaxation behaviour of the metapopulation model introduced in section \ref{sec:model}.
\begin{figure}[htp]
\centering
   \includegraphics[width=0.4\columnwidth,height=0.4\columnwidth]{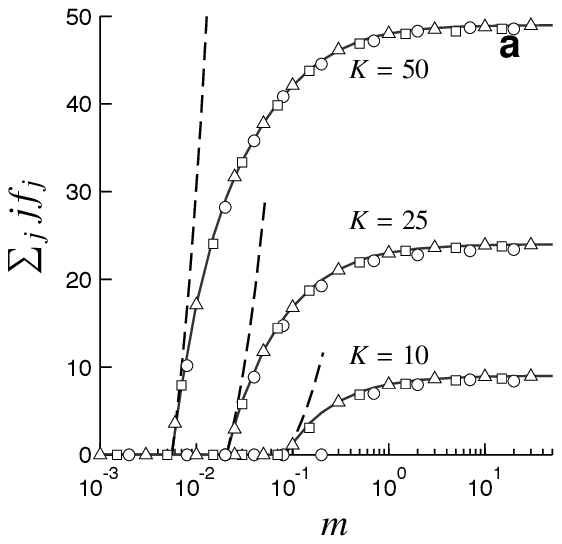}\qquad
   \includegraphics[width=0.4\columnwidth,height=0.4\columnwidth]{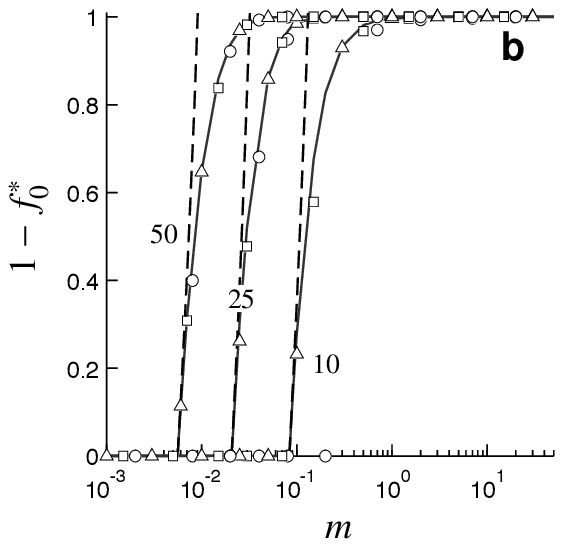}
   \caption{\label{fig:avpopsize_m} (\textsf{\textbf{a}}) Average number of individuals per patch $\sum_{j=1}^{\infty}j f_j$ in the quasi-steady state
    as a function of the emigration rate $m$ for three different values of the carrying capacity $K$.
    Shown are numerical solutions of Eqs.~(\ref{eq:det},\ref{eq:v}), solid lines. Labels indicate the corresponding values of $K$. 
  The growth rate is $r=1.05$.
   Curves corresponding to the asymptotic expression (\ref{eq:mu}), valid
   as $m$ approaches the critical migration rate $m_{\rm c}$,  are shown as dashed lines. Also shown are results of direct numerical simulations
   as described in section \ref{sec:nexp} for metapopulations consisting of $N=1000$ ($\triangle$), $N=100$ ($\square$), and $N=50$ ($\bigcirc$) patches.
   (\textsf{\textbf{b}}) Same but the fraction of occupied patches $1-f_0^\ast$ as a function of $m$. The asymptotic behaviour
   as $m\rightarrow m_{\rm c}$ is given by Eq.~(\ref{eq:f0}) (dashed lines).
   The results for the direct simulations are obtained as follows. For each set of parameters, $30$ stochastic runs up 
to a time smaller than the expected time to extinction are performed (if extinction 
occurs, the simulation is discarded). After the initial transient, $20$ samples are 
taken from each simulation (separated by a time long enough so that the samples are uncorrelated). 
If the simulations consistently become 
extinct during the initial transient we conclude that there is no quasi-steady state, and $f_0^\ast$ is set to unity.}
\end{figure}

%----------------------------------------------------------------------------------------------------------------
\subsubsection{Metapopulation persistence in the limit of infinitely many patches ($N\rightarrow\infty$)} \label{sec:persist}
%----------------------------------------------------------------------------------------------------------------
For sufficiently large emigration rates $m$, the deterministic dynamics (\ref{eq:det},\ref{eq:v}) has
two viable steady states. The state $\ve f=\ven 0$ is unstable, and there is a second steady state $\ve f^\ast$ given by
\begin{align}
f_j^\ast &= f_0^\ast\,\prod_{k=1}^j\frac{b_{k-1}+I^\ast }{\hspace*{4mm}d_k+m_k\hspace*{2mm}}\nn\\
	     &= f_0^\ast \left(\frac{r K}{r-\mu}\right)^{{j}} \frac{\Gamma({j} + I^\ast/r) \, \Gamma(1+ z)}{\Gamma(I^\ast/r) \, \Gamma({j}+1) \, \Gamma({j}+1+z)}\,,
\label{eq:fast}
\end{align}
where $\Gamma(x)$ is the Gamma function, $I^\ast$ is the rate of immigration into a patch in the steady state, 
and $z = \mu K/(r-\mu)$.  Eq. (\ref{eq:fast}) is most easily understood by recognising that
the deterministic dynamics (\ref{eq:det},\ref{eq:v}) takes the form of a master equation for the probabilities $f_j$ that a patch is
occupied by $j$ individuals.  Eq. (\ref{eq:fast}) 
is equivalent to Eqs.~(5a,b) in \citep{Nachman:2000} and to Eqs.~(5,6) in \citep{Casagrandi2002}. In Eq.~(\ref{eq:fast}), the factor $f_0^\ast$
is a normalisation factor (equal to the frequency of empty patches in the steady state) determined
by the requirement that $f_0^\ast+\sum_{j=1}^\infty f_j^\ast = 1$. 
Note that the factors in the product in Eq.~(\ref{eq:fast}) depend upon 
the rate $I^\ast$.
This gives rise to a self-consistency condition for the rate of immigration $I^\ast$ in the steady state:
\begin{equation}
\label{eq:sc}
I^\ast = \sum_{j=1}^\infty m_j f_j^\ast\,.
\end{equation}

The steady state (\ref{eq:fast}) is stable provided the eigenvalues of ${\ma A}$ (this matrix is introduced
in the previous section and its elements are given in appendix \ref{app:A}) have negative real parts. 
This is the case when the steady-state immigration rate $I^\ast$ is larger than zero.
The steady-state immigration rate is expected to decrease when the emigration rate decreases. 
Consider for instance decreasing $m$, defined in Eq. (\ref{eq:rates2}), while keeping all other parameters constant.  
As $m$ approaches a critical value, $m_{\rm c}$, we observe that $I^\ast\rightarrow 0$.  At the same time all components of $\ve f^\ast$ tend to zero, 
and $f_0^\ast \rightarrow 1$.
Below this critical point, that is for $m < m_{\rm c}$, 
the stable steady state ceases to exist (and $\ve f=\ven 0$ turns stable). 
In the limit of infinitely many patches ($N\rightarrow \infty$), the persistence condition 
\begin{equation}
\label{eq:mcc}
m > m_{\rm c}
\end{equation} 
ensures that a stable steady state exists, with $f_j^\ast >0$ for some non-zero values of $j$. 
This persistence criterion for infinitely large metapopulations is
equivalent to the criterion suggested by \citet{Chesson1984} and re-derived by \citet{Casagrandi2002},
namely that the metapopulation persists provided that the expected number of emigrants
from a patch with initially one individual and into which immigration is excluded is greater
than unity. In a slightly different form this principle is quoted by \citet{Hanski98}, namely
that the expected number of successful colonisations out of a given patch during its lifetime in an otherwise
empty metapopulation should be larger than unity. 
We emphasise that this criterion (or any of the equivalent criteria, such as (\ref{eq:mcc})) cannot be used to
determine how stochastic fluctuations in finite metapopulations affect their persistence.

The critical value $m_{\rm c}$ is obtained by analysing the steady-state condition (\ref{eq:sc})
for $m$ close to $m_{\rm c}$. 
We write 
\begin{equation}
\label{eq:defdelta}
m= m_{\rm c}(1+\delta)
\end{equation}
and expand the
steady-state immigration rate in powers of $\delta$ (note that $I^\ast$ vanishes at $\delta=0$):
\begin{equation}
\label{eq:Iexpand}
I^\ast = I_1 \delta + I_2 \delta^2 + \ldots.
\end{equation}
The constant $I_1$ is given in appendix \ref{app:B}.
Expanding Eq.~(\ref{eq:sc}), we find to lowest order in $\delta$ a condition for  
$m_{\rm c}$:
\begin{equation}
\label{eq:mc}
d_1 = m_{\rm c} \sum_{j=2}^\infty j \prod_{k=2}^j\frac{b_{k-1}}{d_k+m_{\rm c} k}\,.
\end{equation}
Fig. \ref{fig:mc} shows how the solution $m_{\rm c}$ of Eq.~(\ref{eq:mc}) depends upon the carrying capacity $K$
for the model introduced in section \ref{sec:model}. The critical migration
rate is shown as a solid line in the left panel of Fig.~\ref{fig:mc}, in the following
referred to as the \lq critical line'.
Above this line, a metapopulation consisting of an infinite number
of patches persists. Expanding the condition (\ref{eq:mc}) for large
carrying capacities $K$ we find that (see appendix \ref{app:C}):
\begin{equation}
\label{eq:mcexpand}
m_{\rm c} \sim  r \sqrt{\frac{r-1}{2\pi K}} 
{\rm exp}\bigg[-K\bigg(1-\frac{\log r}{r-1}\bigg)\bigg]\,.
\end{equation}
Fig.~\ref{fig:mc} also shows $f^\ast_j$ as a function
of $j$ for six  stable steady states corresponding to different
values of $m$ and $K$. Results similar to those shown in the six panels on the right-hand side
of Fig.~\ref{fig:mc} are given in 
Fig.~2 in \citep{Nachman:2000}, and in Fig.~2 {\bf a}-{\bf c} in \citep{Casagrandi2002}.

One observes how the number of empty patches $f_0^\ast$ tends towards unity as the critical line is approached.
To leading order in $\delta$ we have:
\begin{equation}
\label{eq:f0}
f_0^\ast = 1-c_1 \delta\,.
\end{equation}
The constant $c_1$ is given in appendix \ref{app:B}. 
Similarly, the expected number of individuals per patch approaches zero:
\begin{equation}
\label{eq:mu}
{\sum_{j=1}^\infty j f_j^\ast} \sim c_2 \delta
\end{equation}
as $\delta\rightarrow 0$.
The constant $c_2$ is given in appendix \ref{app:B}.  
Fig.~\ref{fig:avpopsize_m} shows
the average number of individuals per patch, and $1-f_0^\ast$ as a function of $m$.  Shown are 
numerical solutions of the steady-state condition (\ref{eq:sc}), solid lines, of Eqs. (\ref{eq:f0}) and (\ref{eq:mu}),
dashed lines, and of direct numerical simulations, symbols, of the metapopulation model described 
in section \ref{sec:nexp}. The numerical experiments were performed for $N=50, 100$ and $1000$ 
patches and averaged over an ensemble of different realisations.
In principle, in a finite metapopulation
no stable steady states exist with $f_j^\ast>0$ for some $j\geq 1$. But the larger  the number of patches,
the larger the average time to extinction is expected to be.  $N=1000$, it turns out, is sufficiently large 
for the parameter values chosen that extinction did not occur during the simulations. 
Consequently we observe good agreement between the direct numerical simulations and the numerical solution
of the deterministic steady-state condition.  When the number of patches is small, by contrast, this is no longer the case.
In order to determine the quasi-steady state in such cases, the simulations must be run for a time large enough for the initial transient to die out, 
but shorter than the expected time to extinction.
Stochastic realisations leading to extinction during the simulation time must be discarded.

\begin{figure}[htp]
\centering
   \includegraphics[width=0.4\columnwidth,height=0.4\columnwidth]{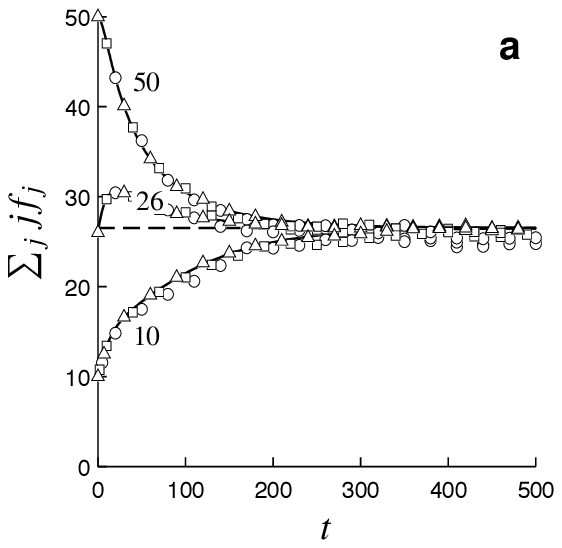}\quad
   \includegraphics[width=0.4\columnwidth,height=0.4\columnwidth]{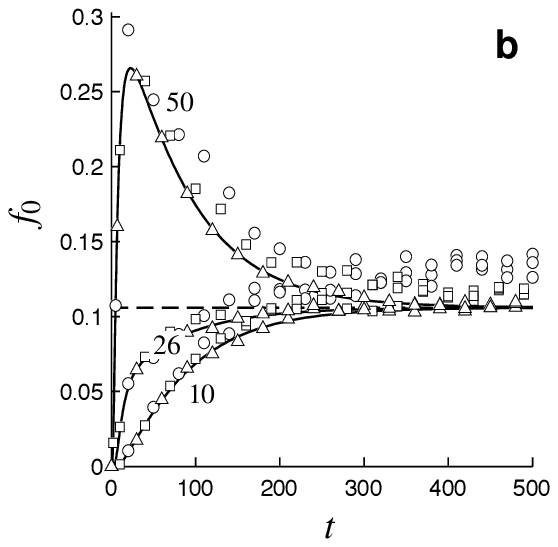}
   \caption{\label{fig:avpopsize_t} (\textsf{\textbf{a}}) Average number of individuals per patch $\sum_{j=1}^{\infty}j f_j$ as a function of time for $r=1.05$, $K=50$, and $m=0.0167$.  
Shown are numerical solutions of Eqs.~(\ref{eq:det},\ref{eq:v}) for three different initial conditions:
   ten individuals per patch, $26$ individuals per patch, and $50$ individuals per patch as indicated by labels in the figure.
   Also shown are results of direct simulations as described in section \ref{sec:nexp},
   for $N=1000$ ($\triangle$), $N=100$ ($\square$) and $N=50$ ($\bigcirc$), for each of the three initial conditions.
   The steady-state value is shown as a dashed line.
   (\textsf{\textbf{b}}) Same, but the frequency of empty patches $f_0$ as a function of time.
   Each point of the direct simulations corresponds to an average over 100 stochastic realisations conditional on no extinction (no extinction actually occurred during these simulations).}
\end{figure}

%----------------------------------------------------------------------------------------------------------------
\subsubsection{Relaxation dynamics in the limit of  $N\rightarrow\infty$ and Levins' model}
%----------------------------------------------------------------------------------------------------------------
We continue to discuss metapopulations with infinitely many patches. In this section
we analyse how the metapopulation relaxes to the stable steady state $\ve f^\ast$.
This question is important for two reasons. First, when the relaxation time
is much smaller than the expected time to extinction (that is, when the number of patches
is large enough), Eq.~(\ref{eq:det}) describes the relaxation dynamics well.
Second, and more importantly, the deterministic dynamics exhibits a \lq slow mode' in
the vicinity of the critical line (in Fig.~\ref{fig:mc}): for small values
of $\delta$, it turns out, there is a particular linear combination of the variables
$f_j$ that relaxes slowly, with a rate proportional to $\delta$. This slow
mode dominates the relaxation dynamics which is thus essentially one-dimensional
for small values of $\delta$. This fact is well known in the theory of dynamical systems
\citep{GuH83}, see also \citet{Dyk94}. For our model, this fact has important implications. Essentially,
the deterministic dynamics close to the critical line in Fig. \ref{fig:mc} reduces to Levins' model, 
as we demonstrate in this section.

Fig.~\ref{fig:avpopsize_t} shows the relaxation of the average number of individuals
and of the frequency of empty patches $f_0$ to their values in the stable steady state $\ve f^\ast$,
for three different initial conditions. Shown are numerical solutions of Eqs.~(\ref{eq:det},\ref{eq:v}),
solid lines, as well as results of direct numerical simulations, symbols. The numerical
experiments were performed for $N=50$, $100$ and $1000$ patches, and averaged over $100$ realisations.  
Even for $50$ patches, and for the parameter values chosen in Fig.~\ref{fig:avpopsize_t}, 
the average time to extinction is large enough so that extinction did not occur in the numerical simulations. 
The agreement between the direct numerical simulations for $N=1000$ and the solution
of Eqs.~(\ref{eq:det},\ref{eq:v}) is good. For $N=50$ and $100$ we can observe deviations to the solution of 
Eqs.~(\ref{eq:det},\ref{eq:v}). Here the effect of the finite number $N$ of patches becomes apparent.

\begin{figure}[htp]
\centering
   \includegraphics[width=0.4\columnwidth,height=0.4\columnwidth]{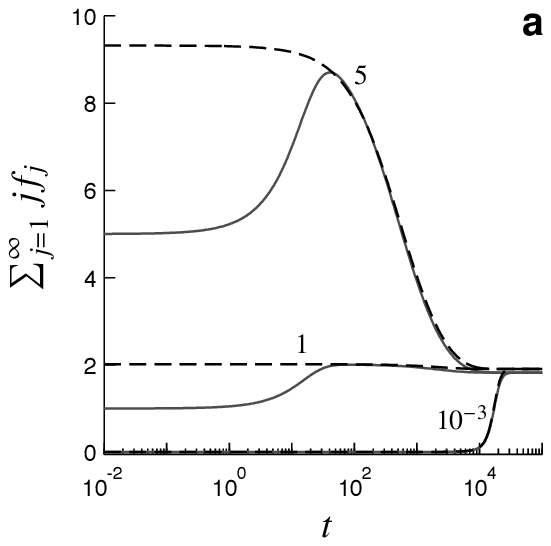}\qquad
   \includegraphics[width=0.4\columnwidth,height=0.4\columnwidth]{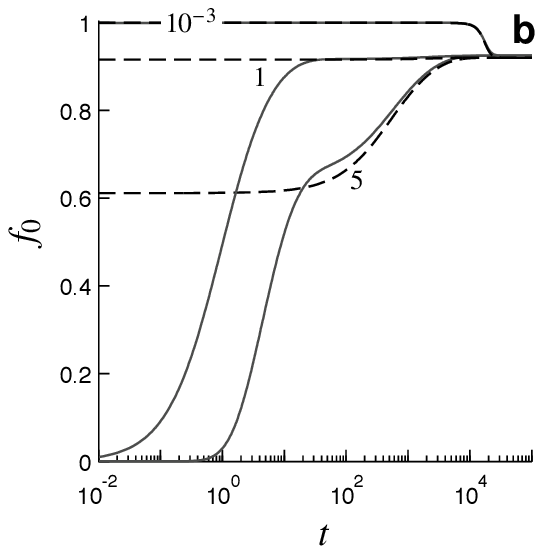}
   \caption{\label{fig:avpopsize_slow} Deterministic relaxation towards the stable steady state $\ve f^\ast$.
   (\textsf{\textbf{a}}) Average number of individuals per patch, $\sum_{j=1}^\infty jf_j$, as a function of time for $r=1.05$, $K=50$, and $\delta=0.05$.
   Shown as solid lines are numerical solutions of Eqs.~(\ref{eq:det},\ref{eq:v}) for three different initial conditions 
   (indicated by labels in the figure):
   five individuals per patch, one individual per patch,  and $10^{-3}$ individuals per patch on average.
   Also shown are solutions of Eqs.~(\ref{eq:Qa}) and (\ref{eq:Le1}) (dashed lines).
   (\textsf{\textbf{b}}) Same, but the frequency of empty patches $f_0$ as a function of time. }
\end{figure}

In general Eqs.~(\ref{eq:det},\ref{eq:v}) must be solved numerically. We now
show that the deterministic dynamics simplifies considerably close to the critical line in Fig.~\ref{fig:mc},
when $\delta$ is small. At $\delta=0$ we have that $I^\ast=0$, $f_0^\ast=1$ and
consequently $\ve f^\ast=\ven 0$. For small positive values of $\delta$,
the components of $\ve f^\ast$ are of order $\delta$, and we expand Eq.~(\ref{eq:det}) in powers of  $\delta$ and $\ve f$
\citep{Dyk94}:
\begin{equation}
\label{eq:expand}
v_i = \sum_j A_{ij}^{(0)} f_j + m_{\rm c}  \delta \sum_j A_{ij}^{(1)}  f_j  + \frac{1}{2} m_{\rm c}(1+\delta)
\sum_{jk} A_{ijk}^{(2)}  f_j f_k\,.
\end{equation}
Here $A_{ij}^{(0)}$ are the elements of the matrix ${\ma A}$ evaluated at the critical line ($\delta=0$).
They are obtained from (\ref{eq:A}):
\begin{alignat}{2}
\label{eq:Ac}
A_{ij}^{(0)} ={}& b_{i-1} \delta_{i-1j} + (d_{i+1}+m_{\rm c}(i+1)) \delta_{i+1j} &&\\ 
&-(b_i+ d_i+m_{\rm c}i)\delta_{ij}  
&&\quad \mbox{for $i> 1$}\,, \nn \\
A_{1j}^{(0)}= {}& (d_{2}+2m_{\rm c}) \delta_{j2} -(b_1+d_1+m_{\rm c})\delta_{j1} + m_c j\,. && \quad \mbox{}\,\nn
\end{alignat}
The slow mode exists because this matrix has one zero eigenvalue, $\lambda_1^{{(0)}}$.
All other eigenvalues are negative. At finite but small values of $\delta$
the corresponding matrix (\ref{eq:A}) evaluated at the steady state ${\ve f}^\ast$ has
one small eigenvalue, $\lambda_1 \sim -\delta$. 
To identify the slow mode we diagonalise ${\ma A}^{(0)}$ given by (\ref{eq:Ac}).  
Since this matrix is not symmetric, its left and right eigenvectors differ:
\begin{equation}
\label{eq:lambda}
{\ma A}^{(0)} \ve R_\alpha = \lambda_\alpha^{{(0)}} \ve R_\alpha\,,\quad 
\ve L_\beta\tr \,{\ma A}^{(0)} = \ve L_\beta\tr \,\lambda_\beta^{{(0)}}\,,\quad
\mbox{for $\alpha,\beta = 1,2,\ldots$}.
\end{equation}
Here $\ve L^{\sf T}_\alpha=(L_{1\alpha},L_{2\alpha},\ldots)$ denotes a left eigenvector
with components $L_{1\alpha},L_{2\alpha},\ldots$. $\ve R_\alpha$ denotes
a right eigenvector with components $R_{1\alpha},R_{2\alpha},\ldots$.
We take the eigenvectors to be bi-orthonormal, 
$\ve L_\beta\tr \ve R_\alpha = \delta_{\alpha\beta}$. As before, the eigenvalues
are ordered as $0 \geq \lambda_1^{(0)} > \lambda_2^{(0)} > \ldots$. 
Multiplying Eq.~(\ref{eq:expand}) from the left with $\ve L_\alpha \tr$ and
making use of Eq.~(\ref{eq:det}) yields the equations of motion of
 $Q_\alpha = \ve L_\alpha \tr \ve f$ (that is, the equations governing the time-evolution of $Q_\alpha$):
\begin{align}
\label{eq:expand2}\frac{{\rm d}Q_\alpha}{{\rm d}t} ={}&  \ve L_\alpha \tr {\ma A}^{(0)}
 \ve f + m_{\rm c}\delta \sum_{ij\beta} L_{\alpha i} A_{ij}^{(1)} R_{\beta j} Q_\beta\\ 
&+\frac{1}{2}  m_{\rm c} (1+\delta) 
\sum_{ijk\mu\nu} L_{\alpha i} A_{ijk}^{(2)}  R_{\mu j} R_{\nu k} Q_\mu Q_\nu\,.\nn
\end{align}
At $m = m_{\rm c}$ (that is, for $\delta=0$) we have that $\lambda_1 = \lambda_1^{(0)} = 0$.  For small values of $\delta$ we find 
$|\lambda_1| \ll |\lambda_\alpha|$ for $\alpha> 1$. 
While $\lambda_\alpha$, $\alpha > 1$, approach constants as $\delta \rightarrow 0$, $\lambda_1 = \ve L_1 \tr {\ma A} \ve R_1 \propto -\delta$ tends to zero in this limit.
This implies
\begin{equation}
\Big| \frac{{\rm d}Q_1}{{\rm d}t}\Big| \ll \Big|\frac{{\rm d}Q_\alpha}{{\rm d}t}\Big|\quad
\mbox{for $\alpha > 1$.}
\end{equation}
In the following, $Q_1$ is therefore referred to as \lq slow mode', whereas
$Q_{\alpha}$ for $\alpha > 1$ are termed fast variables. 
As explained by \citet{Dyk94}, the fast variables rapidly approach quasi-steady states
\begin{equation}
\label{eq:Qa}
 Q_\alpha \approx -\frac{a_{\alpha 1}\delta Q_1 + a_{\alpha2}  Q_1^2}{\lambda_\alpha^{{(0)}}}\quad\mbox{for $\alpha > 1$}
\end{equation}
that depend on the instantaneous value of the slow variable, $Q_1$. Terms including fast variables ($Q_\alpha$ for $\alpha >1$)
are not kept on the right-hand side of Eq.~(\ref{eq:Qa}) because they are of higher order in $\delta$. This is a consequence
of the fact that close to the steady state, $Q_1$ is of order $\delta$. Eq.~(\ref{eq:Qa})  is correct to order $\delta^2$.
The coefficients $a_{\alpha 1}$ and $a_{\alpha2}$ are given by
\begin{align}
\label{eq:c11}
a_{\alpha 1}={}&  m_{\rm c}  \sum_{ij} L_{\alpha i} A_{ij}^{(1)} R_{1j} \,,\\
a_{\alpha 2}={}&\frac{m_{\rm c}}{2} \sum_{ijk} L_{\alpha i} A_{ijk}^{(2)} R_{1j} R_{1k} \,,\nn
\end{align}
for $\alpha = 1,2,\ldots$. The elements $A_{ij}^{(1)}$ and   $A_{ijk}^{(2)}$ are given in appendix \ref{app:A}.
The equation of motion for the slow mode $Q_1$  is to third order in $\delta$:
\begin{align}
\label{eq:Le0}
\frac{{\rm d}Q_1}{{\rm d}t}  ={}& a_{11} \delta Q_1 + a_{12}(1+\delta) Q_1^2\\
\nonumber &+ \mbox{ terms of order $\delta^3$ involving $Q_\beta$ for $\beta > 1$}\\
\nonumber &+ \mbox{ terms of higher order in $\delta$.}
\end{align}
In the remainder of this subsection we neglect the $\delta^3$-terms and higher-order terms in Eq.~(\ref{eq:Le0}).
The $\delta^3$-terms are discussed in subsection \ref{sec:pext}. To second order in $\delta$, the equation
of motion for $Q_1$ is:
\begin{align}
\label{eq:Le1}
\frac{{\rm d}Q_1}{{\rm d}t}  = a_{11} \delta Q_1 + a_{12} Q_1^2\,.
\end{align}
According to Eq.~(\ref{eq:c11}), the coefficients $a_{11}$ and $a_{12}$ are given in terms
of the components $L_{1j}$ of the left eigenvector $\ve L_1$ and the components $R_{1j}$ of the right eigenvector $\ve R_1$
of ${\ma A}^{(0)}$.
For the right eigenvector we find, from Eqs.~(\ref{eq:Ac}) and (\ref{eq:lambda}),
\begin{equation}
\label{eq:R}
R_{1j} = R_{11} \prod_{k=2}^j \frac{b_{k-1}}{d_k+m_{\rm c}k}\quad\mbox{for $j>1$}\,,
\end{equation}
the first component being $R_{11}$. For the left eigenvector, we obtain the following recursion (with the boundary condition $L_{1j}=0$ for $j=-1$):
\begin{equation}
\label{eq:Lrec}
L_{1j+1}-L_{1j} = \frac{d_j+m_{\rm c}j}{b_j} (L_{1j}-L_{1j-1}) -\frac{m_{\rm c}j}{b_j} L_{11}\,.
\end{equation}
This recursion is solved by:
\begin{align}
\label{eq:L}
&{}L_{11}\,,\quad L_{12}= L_{11}\left( 1+\frac{d_1}{b_1}\right)\,,\quad\mbox{and}\quad\\
&{}L_{1j} = L_{11} \left( 1+\frac{d_1}{b_1} + \frac{m_{\rm c}}{b_1} \sum_{n=2}^{j-1} \,\,\sum_{k=n+1}^\infty \frac{k R_{1k}}{n R_{1n}}\right)
\quad\mbox{for $j>2$}\,.\nn
\end{align}
We show in  appendix \ref{app:C} that the elements of $\ve L_1 \tr$ approach the following limiting
form as $K\rightarrow \infty$:
\begin{equation}
\label{eq:LlargeK}
L_{11}\,, \quad L_{12} = L_{11}\, \frac{r+1}{r}\,,\quad\mbox{and}\quad L_{1j}= L_{11}\, r\frac{1-r^{-j}}{r-1}\,.
\end{equation}
We choose $L_{11}$ such that  $L_{1j}$ approaches unity for $j\gg 1$ as $K\rightarrow\infty$.
This corresponds to the choice $L_{11} = (r-1)/r$, resulting in
\begin{equation}\label{eq:L1j}
L_{1j} = 1-r^{-j}
\end{equation}
in the limit of $K\rightarrow \infty$.  In this limit, and for large values of $r$, we see that $\ve L_1 \tr$ approaches the vector $(1,1,\ldots)$.
The convention leading to Eq.~(\ref{eq:L1j}) also fixes $R_{11}$ which must be chosen so that $\ve L_1 \tr \ve R_1 =1$.
Explicit expressions for $a_{11}$ and $a_{12}$, obtained from Eqs.~(\ref{eq:c11}), (\ref{eq:R}), and (\ref{eq:L})  
are given in appendix \ref{app:B}.

 Eq. (\ref{eq:Le1}) has two steady states $Q_1=0$ and $Q_1^\ast = -a_{11}\delta/a_{12}$.
 Note that $Q_1^\ast$ is positive since $a_{12} < 0$.    
 These two steady states correspond to
 the steady states $\ve f=0$ and $\ve f^\ast$ of Eq.~(\ref{eq:det}).
 Comparison with Eq.~(\ref{eq:fast}) shows that $\ve R_1$ is proportional to $\ve f^\ast$
 to lowest order in $I^\ast$ (that is, to lowest order in $\delta$). In fact we have
 \begin{equation}
\label{eq:fR}
\ve f^\ast = \ve R_1  Q_1^\ast\quad+\quad\mbox{higher orders in $\delta$}\,.
\end{equation}

Fig.~\ref{fig:avpopsize_slow} illustrates how the deterministic dynamics relaxes to the stable steady state $\ve f^\ast$.
Shown are the expected number of individuals per patch, $\sum_j jf_j$ and the fraction of empty patches as functions
of time, determined from numerical solutions of Eqs.~(\ref{eq:det},\ref{eq:v}). Curves for three different initial conditions are shown (solid lines).
The corresponding solutions of Eq.~(\ref{eq:Le1}) are shown as dashed lines.
Initially the variable $Q_1$ is not much slower than the $Q_\alpha$-variables for $\alpha > 1$,
and the approximate one-dimensional dynamics (\ref{eq:Le1}) is not a good approximation. But
the solution of Eqs.~(\ref{eq:det},\ref{eq:v}) rapidly relaxes to a form where $Q_1$ becomes
a slow mode, and Eq.~(\ref{eq:Le1}) accurately describes the slow approach to the steady state.

Comparing Eqs.~(\ref{eq:Le1}) and (\ref{eq:LevinsModel}) shows that the slow mode $Q_1$
obeys Levins' equation. The results of this subsection allow us to compute $e$ and $c$
in terms of $r$, $K$, and $\delta$. We find to order $\delta$
\begin{align}
\label{eq:Levin_ce}
c-e ={}& a_{11} \delta \,,\\
c ={}&  -a_{12}+\quad\mbox{terms of order $\delta$ }\,.
\nonumber
\end{align} 
The contributions to $c$ of order $\delta$ can be computed explicitly
using the formulae derived above. For the sake of brevity we do not
specify these contributions here. In Sec.~\ref{sec:text} 
we give an explicit formula valid in the limit of large $K$.

\begin{figure}[htp]
\centering
   \includegraphics[width=0.4\columnwidth,height=0.4\columnwidth]{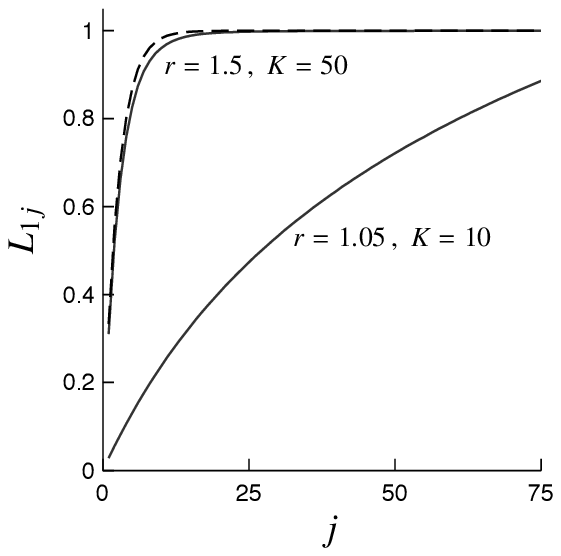}
      \caption{\label{fig:L1} Shows the components $L_{1j}$ of the left eigenvector $\ve L_1$ (see Eq.~\eqref{eq:lambda}) as a function of $j$, for $r=1.5$, $K=50$,
    and for $r=1.05$, $K=10$ (solid lines). The dashed line corresponds to the limit $K \to \infty$, Eq.~\eqref{eq:L1j}, evaluated for $r=1.5$. }
\end{figure}

We see that that Levins' model describes the dynamics
of the metapopulation regardless of whether the patches are strongly coupled or not, 
provided the metapopulation is sufficiently close to criticality (that is, close to
the critical line in Fig.~\ref{fig:mc}). But we emphasise that in general
the slow variable is not the fraction of occupied patches as envisaged by Levins. 
The variable $Q_1$ is given by $\ve L_1 \tr \ve f $.
Fig.~\ref{fig:L1} shows how the components $L_{1j}$ of $\ve L_1$ depend
upon $j$. When the local population dynamics is
fast compared to the migration dynamics (for $r=1.5$ and $K=50$), the vector is approximately given by $(1,1,1,\ldots)$, see also Eq.~\eqref{eq:L1j}.
In this case, $Q_1$ is approximately equal to the fraction $Q$ of occupied patches. This is the limit of
time-scale separation considered by Levins. We have thus derived the coefficients
appearing in Eq.~(\ref{eq:LevinsModel}) from a stochastic, individual-based metapopulation model defined
in terms of the life history of its inhabitants (given by local birth and death rates, as well as the emigration rate).
The asymptotic behaviours of $a_{11}$ and $a_{12}$ for large values of $K$ are given by (see appendix \ref{app:C}):
\begin{equation}
\label{eq:a11_a12_asy}
a_{11} \sim -a_{12} \sim  \sqrt{\frac{K(r-1)^3}{2\pi }}
\exp\Big[-K\Big(1-\frac{\log r}{r-1}\Big)\Big]\,.
\end{equation}
When the patches are strongly mixed by migration, then the interpretation of $Q_1$ is different.
Fig.~\ref{fig:L1} shows that for $r=1.05$ and $K=10$, the components $L_{1j}$ are
roughly proportional to $j$ in the relevant range of $j$ (where $f_j^*$ is not too small). 
In this case, therefore, the slow mode is interpreted as the average number of individuals
per patch, which in turn is proportional to the immigration rate, Eq.~(\ref{eq:I1}).

In summary, in this section we have shown how the deterministic metapopulation
dynamics simplifies in the vicinity of the bifurcation (critical line in Fig. \ref{fig:mc}), that
is, for small values of $\delta$. In this limit, the deterministic dynamics
is essentially one-dimensional, and of the same form as the deterministic
equation for the fraction of occupied patches originally suggested by \citet{Levins1969}.
Commonly it is argued that the form of Eq.~(\ref{eq:LevinsModel}) is appropriate
when the local dynamics is much faster than migration. We have seen
that this is not a necessary condition. In general, Eq. (\ref{eq:LevinsModel}) 
describes the deterministic dynamics of metapopulations on the brink of
extinction (for small values of $\delta$). We find that the variable $Q$
is indeed given by the fraction of occupied patches when the local-patch dynamics
is fast. In general, however, $Q$ has a different interpretation.
Finally, we have been able to relate the rates $c$ and $e$ to the parameters of the individual-based, stochastic model
(namely $r$, $K$, and $\delta = (m - m_{\rm c}) / m_{\rm c}$). 

Last but not least we emphasise that  patch extinction in our model is entirely due to demographic
fluctuations. It is possible to generalise our model so that the extinction rate accounts, in addition, 
for environmental stochasticity, possibly including \lq killing'
\citep{coolen_schrijner_2006}: transitions from an arbitrary state of a patch directly to extinction of that patch, local catastrophes in other words. 

%================================================================================================================
\subsection{Finite number of patches}
\label{sec:pext}
%================================================================================================================
In the previous section we summarised our results on metapopulation dynamics
in the limit of $N\rightarrow \infty$. The subject of the present section
is the dynamics  for metapopulations consisting of a finite number $N$ of patches.
In this case the fluctuations inherent in the birth-, death- and migration processes
lead to fluctuations around the 
steady state $\ve f^\ast$. When the number
of patches is large, these fluctuations are expected to be small. But they
are essential: in a finite metapopulation the only absorbing state is $\ve f = \ven 0$.
In other words, the fluctuations turn the stable steady state $\ve f^\ast$  of the deterministic 
dynamics into an unstable one.   

%----------------------------------------------------------------------------------------
\subsubsection{Fluctuations around the quasi-steady state}\label{sec:fluct}
%----------------------------------------------------------------------------------------

\begin{figure}[htp]
\centering
	\includegraphics[width=0.4\columnwidth,height=0.4\columnwidth]{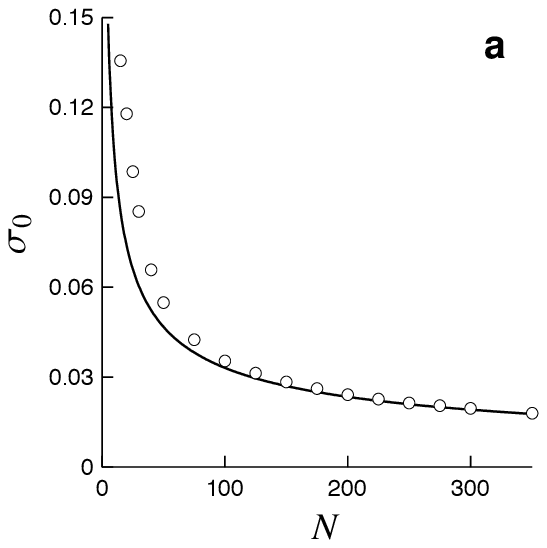} \qquad
	\includegraphics[width=0.4\columnwidth,height=0.4\columnwidth]{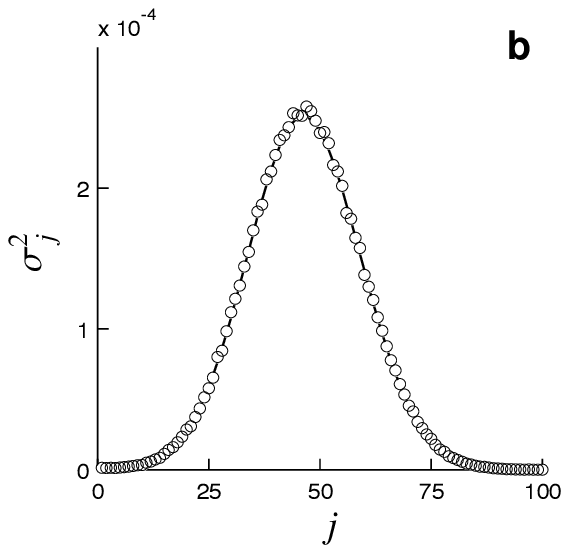}
	\caption{\label{fig:sigma}Size of the fluctuations around the quasi-steady state of the distribution \eqref{eq:ansatz}. (\textsf{\textbf{a}}) Standard deviation $\sigma_0$ of the fraction of empty patches as a function of the number of patches $N$. Comparison of $\sigma_0$, calculated from $\mathbf{C}$ in Eq.~\eqref{eq:C2} and employing Eq.~\eqref{eq:sig0} (solid line), to direct numerical simulations ($\bigcirc$). 
Parameters: $r=1.05$, $K=50$, $m=0.02$. (\textsf{\textbf{b}}) Variances $\sigma_j^2$ of the fraction of patches $f_j$ with $j$ individuals as a function of $j$ ($j = 1,2,\dotsc$). Comparison of $\sigma^2_j$ calculated using Eq.~\eqref{eq:C2} (solid line), to direct numerical simulations ($\bigcirc$) for $N=100$. Parameters: $r=1.5$, $K=50$, $m=2.5 \times 10^{-5}$. }
\end{figure}

For finite values of $N$, the population fluctuates around its quasi-steady state, as pointed out in section \ref{sec:qssfinite}. 
These fluctuations are Gaussian and of order $N^{-1}$  (cf. Eqs.~\eqref{eq:ansatz} and \eqref{eq:S2}).
Fig.~\ref{fig:sigma}a shows how the the standard deviation of the fraction of empty patches, $\sigma_0$, 
depends on the number of patches, $N$.
The standard deviation is calculated from the covariance matrix ${\ma C}$, Eq.~\eqref{eq:C2}, as
\begin{equation}\label{eq:sig0}\begin{split}
	\sigma_0^2 ={}& \mom{f_0^2} - \mom{f_0}^2 
	=\frac{1}{N}\sum_{i,j=1}^{\infty} C_{ij} \, .
\end{split}
\end{equation}
Comparing the analytical approximation of $\sigma_0$, calculated using Eqs.~(\ref{eq:C2},\ref{eq:sig0}), 
with simulations of the full stochastic model, we find good agreement when $N \ge 100$. 
Fig.~\ref{fig:sigma}b shows the variance for each $\sigma_j^2 = \mom{f_j^2} - \mom{f_j}^2$ as a function of $j$, for $N = 100$.

%----------------------------------------------------------------------------------------
\subsubsection{Most likely path to extinction at finite but large values of $N$} \label{sec:optpath}
%----------------------------------------------------------------------------------------
When the number of patches is large, the state $\ve f^\ast$ may be very long-lived (and is thus referred to as a quasi-steady state). The 
methods described in section \ref{sec:exp}
allow us to systematically analyse the properties of this quasi-steady state in terms
of  Eq.~(\ref{eq:HEM}). While the deterministic dynamics in the 
limit of $N\rightarrow\infty$ is given by Eq.~(\ref{eq:det}),
Eq.~(\ref{eq:HEM}) describes the the stochastic fluctuations of the quasi-stable distribution at finite
values of $N$ prior to extinction.
We note that by setting $\ve p=\ven 0$ in
the equations of motion  [Eq.~(\ref{eq:HEM})], the deterministic dynamics  [Eq.~(\ref{eq:det})] is obtained.
In order to describe the fluctuations of a finite metapopulation around $\ve f^\ast$
we need to find the most likely path from $\ve f^\ast$ to a point $\ve f$ in the vicinity,
namely the path from $\ve f^\ast$ to $\ve f$ with extremal action [Eq.~(\ref{eq:action})], satisfying $H(\ve f(t), \ve p(t))=0$,
as well as the boundary condition [Eq.~(\ref{eq:bc0})]. 

The tails of the quasi-steady state distribution
near $\ve f=\ven 0$ are obtained by computing the path
from $\ve f^\ast$ to $\ve f=\ven 0$ with extremal action. This path is termed the most likely path to extinction, 
because in the limit of large values of $N$
the trajectories of the stochastic, individual-based dynamics to extinction are expected
to fluctuate tightly around this path. The most likely path to extinction
leaves the saddle point $(\ve f^\ast,\ven 0)$ along an unstable direction. Therefore this path cannot 
directly connect to the origin $(\ven 0,\ven 0)$.
It turns out that the dynamics (\ref{eq:HEM}) exhibits a saddle point at $\ve f=\ven 0$ and at non-vanishing momenta, 
which we call $\ve p^\ast$.
The steady-state condition ${\rm d}\ve p/{\rm d}t=\ven 0$ gives rise to 
a recursion relation for the momentum components of $\ve p^\ast$ of this saddle point:
\begin{equation}
\label{eq:precursion}
{\rm e}^{p^\ast_{k+1}-p^\ast_k} -1=
\frac{d_k}{b_k}(1-{\rm e}^{-p^\ast_k+p^\ast_{k-1}})
+\frac{m_k}{b_k}(1-{\rm e}^{p^\ast_1-p^\ast_k+p^\ast_{k-1}})\,.
\end{equation}
The problem lies in finding the most likely path from $(\ve f^\ast,\ven 0)$ to $(\ven 0,\ve p^\ast)$.  
The manifold connecting these two points is infinite dimensional. It is straightforward
to truncate the system of equations (\ref{eq:HEM}) at some large value of $j$, but finding
the extremal path in the high-dimensional manifold  (for instance by a numerical
shooting method starting in the vicinity of $\ve f^\ast$) is very difficult.

However, the problem simplifies considerably in vicinity of the critical line in the left panel of Fig.~\ref{fig:mc}.
When $\delta$ is small, the dynamics (\ref{eq:HEM}) is approximately
two-dimensional \citep{Dyk94}, since the linearisation (\ref{eq:J}) has two small eigenvalues
when $\delta \ll 1$. At $\delta = 0$ we write 
\begin{align}
{\bf J}^{{(0)}}\, \bm{\mathcal{R}}_\alpha ={}& \lambda_\alpha^{{(0)}} \bm{\mathcal{R}}_\alpha\label{eq:posevJ}\,,\\
{\bf J}^{{(0)}}\, \bm{\mathcal{R}}^{\prime}_\alpha ={}& -\lambda_\alpha^{{(0)}} \bm{\mathcal{R}}^{\prime}_\alpha \, , \nn
\end{align}
and corresponding equations for the 
left eigenvectors $\bm{\mathcal{L}}^{\sf T}_{\alpha}$ and $\bm{\mathcal{L}}^{\prime \sf T}_\alpha$.
Here ${\bf J}^{(0)}$ denotes the matrix ${\bf J}$ (see Eq.~\eqref{eq:J}) evaluated at $\delta = 0$, and $\lambda_\alpha^{{(0)}}$ are the eigenvalues of ${\ma A}^{(0)}$. 
The left and right eigenvectors
of ${\bf J}^{{(0)}}$ can be written in terms of those of ${\ma A}^{(0)}$:
\begin{equation}\label{eq:calR}
        \ve{\mathcal{R}}_{\alpha}^{\prime}
        = \begin{pmatrix} \ven 0 \\ {\ket{L}}_{\alpha} \end{pmatrix}\,, \quad \text{and} \quad  
       \ve{\mathcal{R}}_{\alpha} 
        = \begin{pmatrix}{\ket{R}}_{\alpha} \\ \ven 0 \end{pmatrix} \, .
\end{equation}
The left eigenvectors are given by
\begin{equation}\label{eq:calL}
        \ve{\mathcal{L}}_{\alpha}^{\prime \sf T} 
         = \begin{pmatrix} \ven 0, & {\ve{R}}_{\alpha}\tr \end{pmatrix}\,, \quad \text{and} \quad  
        \ve{\mathcal{L}}_{\alpha}\tr = \begin{pmatrix}{\ve{L}}_{\alpha}\tr, & {\ven 0} \end{pmatrix} \,.
\end{equation}

\begin{figure}[htp]
\centering
   \includegraphics[width=0.4\columnwidth,height=0.4\columnwidth]{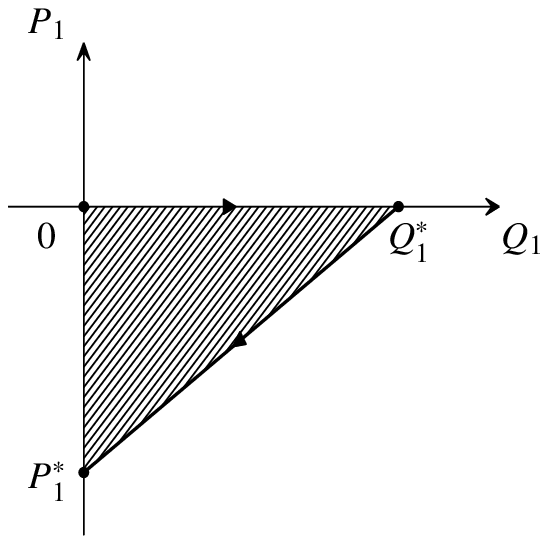}
   \caption{\label{fig:PQ_plane} Illustrates dynamics in $Q_1$-$P_1$ plane, Eq.~(\ref{eq:slowHeq}). 
   See Section \ref{sec:optpath}.
   Similar to the schematic sketch shown in Fig.~\ref{fig:fp_plane}a. 
The path from $(0,0)$ to $(Q_1^*,0)$ corresponds to the deterministic dynamics for $Q_1$, Eq.~\eqref{eq:Le1}. 
The path from the quasi-steady state to the fluctuational extinction point $(P_1^\ast,0)$ is a straight line. 
The action $S$, Eq.~\eqref{eq:action2}, is given by the shaded area: $S=Q_1^\ast P_1^\ast/2$.}
\end{figure}

The slow variables are obtained by projecting $(\ve f,\ve p)^{\sf T}$ onto
$\bm{\mathcal{L}}^{\sf T}_1$ and $\bm{\mathcal{L}}^{\prime \sf T}_1$. They are thus simply given by $Q_1$ and $P_1$, where
\begin{equation}\label{eq:QPalpha}
Q_\alpha = \ve L\tr_\alpha \ve f \quad \mbox{and}\quad P_\alpha = \ve p \tr \ve R_\alpha \,.
\end{equation}
Following \citet{Dyk94}, the slow dynamics of $Q_1$ and $P_1$ is found by expanding the Hamiltonian (\ref{eq:H})
in powers of $\delta$, $\ve f$, and $\ve p$. Starting from Eq.~(\ref{eq:H2}), we use Eq.~(\ref{eq:expand}), 
as well as an expansion of ${\ma D}(\ve f)$ in powers of $\ve f$: noting that ${\ma D}$ vanishes
at criticality, we write
\begin{equation}\label{eq:Bijk}
D_{ij} =  \sum_{k=1}^\infty B_{ijk}^{(1)}   f_k+\ldots\,.
\end{equation}
The elements of $B_{ijk}^{(1)}=\partial D_{ij}/\partial f_k$ are given in appendix \ref{app:A}.
$H(Q_1,P_1)$ has the form  (to third order in $\delta$)
\begin{equation}
\label{eq:slowH}
H(Q_1, P_1) = P_1(a_{11}\delta Q_1 + a_{12} Q_1^2)+ b_{11}Q_1 P_1^2\,.
\end{equation}
The coefficients
\begin{equation}
\label{eq:b11}
b_{11} = \frac{1}{2}\sum_{ijk} L_{1i} B_{ijk}^{(1)} L_{1j} R_{1k}\, ,
\end{equation}
and $a_{11}$, $a_{12}$ are given in appendix \ref{app:B}. The equation of motion for $Q_1$ and $P_1$ is
\begin{equation}
\label{eq:slowHeq}
\frac{{\rm d}Q_1}{{\rm d}t} = \frac{\partial H}{\partial P_1}
\quad\mbox{and}\quad\frac{{\rm d}P_1}{{\rm d}t} = -\frac{\partial H}{\partial Q_1}\,.
\end{equation}
The dynamics determined by Eqs.~(\ref{eq:slowH},\ref{eq:slowHeq}) 
is illustrated in Fig.~\ref{fig:PQ_plane}, similar to Fig.~\ref{fig:fp_plane}.  
The three steady states of Eqs.~(\ref{eq:slowH},\ref{eq:slowHeq}) of interest
for the questions addressed here are:
\begin{align}
\label{eq:s1} &Q_1 = 0\quad \mbox{and} \quad P_1 = 0                               \,,\\
\label{eq:s2} &Q_1 = 0\quad \mbox{and} \quad P=P_1^\ast = -a_{11}\delta/b_{11} \,,\\
\label{eq:s3} &Q_1=Q_1^\ast = -a_{11}\delta/a_{12}\quad \mbox{and} \quad  P_1 = 0\,.
\end{align}
All three steady states are saddle points. The steady state (\ref{eq:s3}) corresponds to the 
saddle point $(\ve f^\ast,\ven 0)$ of the dynamics (\ref{eq:HEM}).
 Note that $a_{12} < 0$.  The steady state (\ref{eq:s2}) corresponds to the saddle point $(\ven 0,\ve p^\ast)$
which lies at the end of the most likely path to extinction. 
In one-dimensional single-step birth-death processes, the corresponding point is commonly
referred to as \lq fluctuational extinction point'. To lowest order in $\delta$ we find for the solution $\ve p^\ast$ of Eq.~\eqref{eq:precursion}
\begin{equation}
\label{eq:pR}
\ve p^\ast = P_1^\ast \ve L_1 \quad+\quad\mbox{higher orders in $\delta$}\, .
\end{equation}
Eq.~(\ref{eq:pR}) furnishes an interpretation
of the left eigenvector $\ve L_1 \tr$ for small values of $\delta$. This vector is proportional to the vector 
$\ve p^\ast$, and the components of this vector define the coordinates of the fluctuational extinction point.
In one-dimensional birth-death processes,
the corresponding value is given by $p^\ast = -\log R_0$ where $R_0$ is the reproductive value. 
An example is the so-called SIS-model \citep{Doering2005,Meerson2010}, a stochastic
model for the duration of the epidemic state of an infectious disease. 
Here the reproductive value $R_0$ is the expected number of infections caused, during its lifetime, by one infected individual 
introduced into a susceptible population. In infinite populations the epidemic persists
provided $R_0>1$. This criterion is precisely analogous to the persistence criteria discussed
in Sec.~\ref{sec:qss}.  When $R_0$ is only slightly larger than unity, $R_0 =1+\epsilon$, say, then $p^\ast \approx -\epsilon$. 

In our case, the vector $\ve L_1 \tr$ plays the role of a reproductive {\em vector} \citep{Fisher30,Samuelson77}.
Expanding the solution of the linearised deterministic dynamics (\ref{eq:det}), in terms
of the eigenvectors of ${\ma A}^{(0)}$ yields
\begin{align}
\delta \ve f(t) = {}& \sum_{\alpha=1}^\infty {\rm e}^{\lambda_\alpha t} \ve R_\alpha \ve L_\alpha\tr \delta \ve f(0) \\
\sim {}&  {\rm e}^{\lambda_1 t} \ve R_1 \ve L_1\tr \delta \ve f(0) \quad\mbox{at large times}\,. \nn
\end{align}
We see that the components $L_{1j}$ of the vector $\ve L_1$ determine how much fluctuations of
the number of patches with $j$ individuals contribute to the relaxation towards the steady state in infinite metapopulations.
In other words, the components of $\ve L_1$ determine the susceptibility of patches
with $j$ individuals to stochastic fluctuations. We shall see below that this susceptibility determines
the average time to extinction.  We note that the correspondence
(\ref{eq:pR}) implies that the solution of the recursion (\ref{eq:precursion}) for $p_j^\ast$
must be equivalent, to lowest order in $\delta$, to the recursion (\ref{eq:Lrec}) for
the components of $\ve L_1$, up to a factor. This is indeed the case.
In analogy with the SIS-model discussed above, the coordinates of the vector $\ve L_1 \tr$ parameterise
the fluctuational extinction point. 

Finally we note that the components of $\ve L_1$ have a simple interpretation in the limit
of $K\rightarrow \infty$. Comparing Eqs.~(\ref{eq:L1j})  and (\ref{eq:ujK}) we see
that the component $L_{1j}$ is given by the probability of a single, isolated patch with
$j$ individuals to eventually reach its carrying capacity, $K$.
This observation concludes our discussion of the nature of the fluctuational extinction point
of Eq.~(\ref{eq:slowHeq}).

The most likely escape path leads from the saddle (\ref{eq:s3}) to
the fluctuation extinction point (\ref{eq:s2}). The path is parameterised by $Q_1$ and $P_1$. Solving $H=0$ for $P_1$
we find from Eq.~(\ref{eq:slowH}) (discarding the trivial solution $Q_1=P_1=0$)
\begin{equation}
\label{eq:instanton}
P_1 = -b_{11}^{-1}(a_{11} \delta + a_{12} Q_1)\,.
\end{equation}
This path corresponds to the straight line from $(Q_1^*,0)$ to $(0,P_1^*)$ in Fig.~\ref{fig:PQ_plane}. 
Eq.~(\ref{eq:instanton}) together with Eqs.~(\ref{eq:fR}) and (\ref{eq:pR}) imply that
the $\ve f$- and $\ve p$-spectra move rigidly towards extinction. In other words, our
analysis shows that on the path to extinction, both $\ve f(t)$ and $\ve p(t)$ retain
their shape (initially given by the right and left eigenvectors $\ve R_1$ and $\ve L_1 \tr$ of ${\ma A}^{(0)}$)
to lowest order in $\delta$.
 
\begin{figure}[htp]
   \centering
   \includegraphics[width=0.4\columnwidth,height=0.4\columnwidth]{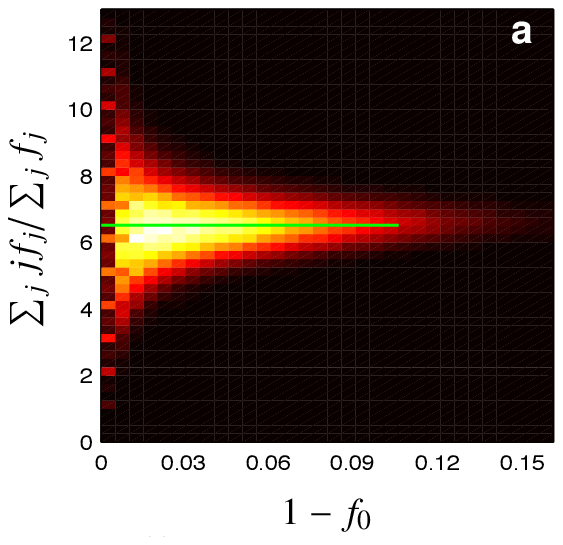}\qquad
   \includegraphics[width=0.4\columnwidth,height=0.4\columnwidth]{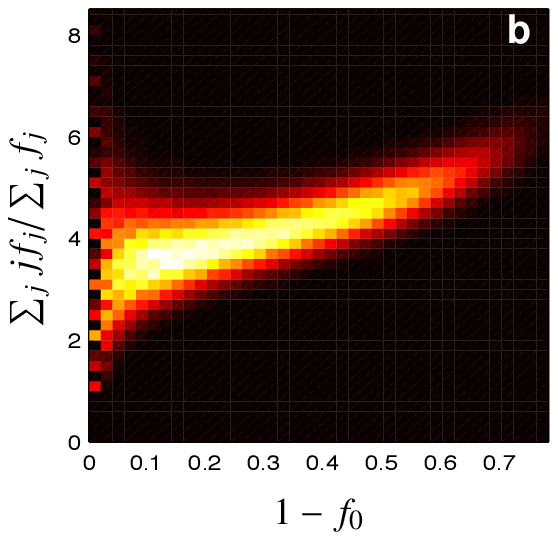}
   \caption{\label{fig:colorprob} Results of direct numerical simulations of the stochastic, individual-based model,
       determining the most likely path to extinction. Shown is the probability distribution of the average number of individuals
       per occupied patch 
      $\sum_{j}jf_j / \sum_{j}f_j$, conditional on the fraction of occupied patches $1-f_0$.
     Parameters: $r=1.5$, $K=10$, $\delta = 0.103$ (corresponding to $m = 0.028$), 
    $N=1000$ (\textsf{\textbf{a}}) and $r=1.05$, $K=10$, $\delta=0.61$ (corresponding to $m = 0.1342$), 
   and $N=250$ (\textsf{\textbf{b}}).
     The probability is colour-coded: high probability corresponds to white, low probability to black. How this
     plot was produced is described in Sec.~\ref{sec:optpath}. Also shown (only in panel \textsf{\textbf{a}}) is the result of the slow dynamics given
by Eqs.~(\ref{eq:fR}), (\ref{eq:pR}), and (\ref{eq:instanton}) (solid green line).}
\end{figure}

In the limit of large carrying capacities (where the critical emigration rate is small)
one might expect that patches become extinct independently of each other. In this limit,
the rate of extinction of single patches in the population is small, and the $\ve f$-spectrum
relaxes rapidly to its rigid shape once a given patch has gone extinct.
It is important to emphasise that the patches are nevertheless coupled
by migration which gives rise, in this limit, to a small rate of colonisation of empty patches.
Thus the average time $T_{\rm ext}$ to extinction for the whole system is not determined by the largest
time of extinction of the $N$ single patches (we discuss the time $T_{\rm ext}$ in the following subsection).
This has strong implications for conservation biology: it indicates the importance of protecting available (possibly empty) patches, even if they are small and prone to extinction.

When the population is strongly mixed by migration, on the other hand, it is surprising
that the $\ve f$- and $\ve p$-spectra move rigidly to extinction.
Migration upholds a balance that causes the shape
of the $\ve f$-spectrum to remain unchanged as the metapopulation
comes closer to extinction. In other words,  just the normalisation  $1-f_0$ changes.

We have attempted to verify the predictions of this subsection 
by direct numerical simulations of the stochastic, individual-based model.
The result is shown in Fig.~\ref{fig:colorprob}. The simulations were performed as follows. For $10,000$ independent
realisations we followed the metapopulation to extinction. For each realisation we analysed
the path to extinction by tracing the stochastic trajectory back in time, starting at the time of extinction.
We traced the trajectories back for the average time it takes to reach, using this procedure,
the number of individuals that corresponds to the quasi-steady state. As a function of time, the average number of individuals
per patch and the fraction of occupied patches was computed. Fig. \ref {fig:colorprob} shows the probability
of observing a given number of individuals in occupied patches, conditional on the fraction of
occupied patches, $1-f_0$. The probability is colour coded: white corresponds to high probability,
black to low probability. The left panel of Fig.~\ref{fig:colorprob} is consistent with the prediction that the $\ve f$-spectrum moves rigidly. This is not the case for the case shown in the right panel of Fig.~\ref{fig:colorprob}. Here $\delta$ is too large,
the two-dimensional approximation to the full Eq.~(\ref{eq:HEM}) is not accurate.
\begin{figure}[htp]   
\centering   
    \includegraphics[width=0.8\columnwidth]{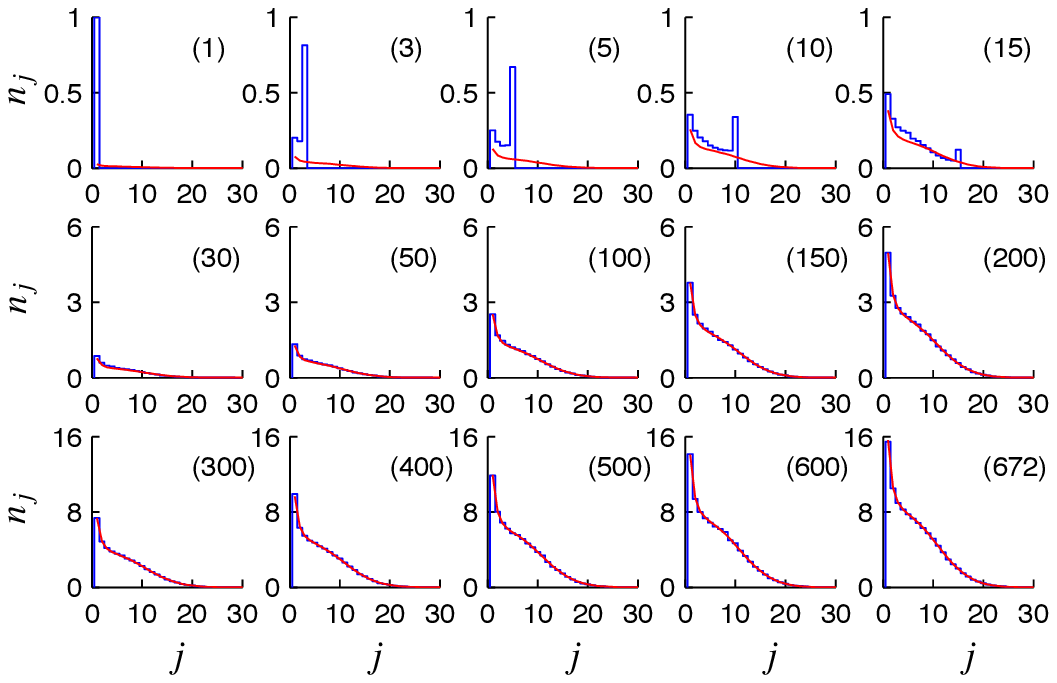}
    \caption{\label{fig:extpath} 
Each panel shows the distribution of the number $n_j$ of patches with $j$ individuals, conditional on that there are  $(n)$ individuals in the metapopulation. Thus the upper, left-most 
panel shows $n_j$ conditional
on that there is one individual, and the lower right-most panel corresponds to $672$ individuals 
(the expected number of individuals in the steady state). 
Results from direct simulations (averaged over $10^4$ stochastic realisations) (blue) are compared to results of numerical integrations of 
Eqs.~(\ref{eq:H},\ref{eq:HEM}) (red). 
Parameters: $r=1.5$, $K=10$, $\delta = 0.103$ ($m = 0.028$), 
$N =1000$. Note that the first value of $j$ in each panel is j=1. In other words, $n_0$ is not shown. }
\end{figure}

An alternative representation of the most likely path to extinction is depicted in Fig.~\ref{fig:extpath}. 
Employing the same simulations described above, Fig.~\ref{fig:extpath} shows the average number $n_j$ of patches with $j$ individuals conditional on the total number of individuals, compared with results obtained by numerically integrating Eq.~\eqref{eq:HEM}. We observe that the agreement is good up to a point 
where $n_j$ becomes too small. When the total number of individuals is small (less than approximately 30 
for the parameters in Fig.~\ref{fig:extpath}) discrete effects start to dominate, and the large-$N$ expansion of the master equation fails to accurately describe the last part of the trajectory towards extinction. 

%----------------------------------------------------------------------------------------------------------------
\subsubsection{Time to extinction for finite but large values of $N$}
\label{sec:text}
\begin{figure}[htp]
\centering
   \includegraphics[width=0.35\columnwidth,height=0.32\columnwidth]{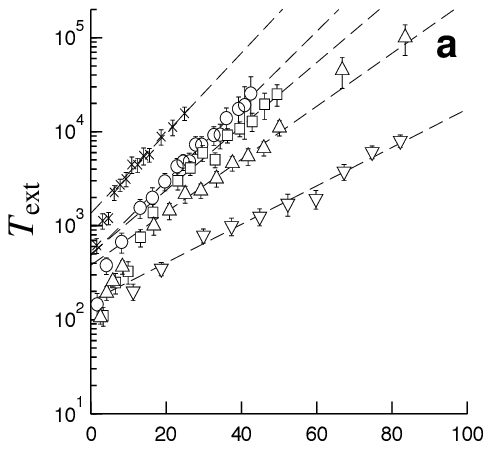}\qquad
   \includegraphics[width=0.35\columnwidth,height=0.32\columnwidth]{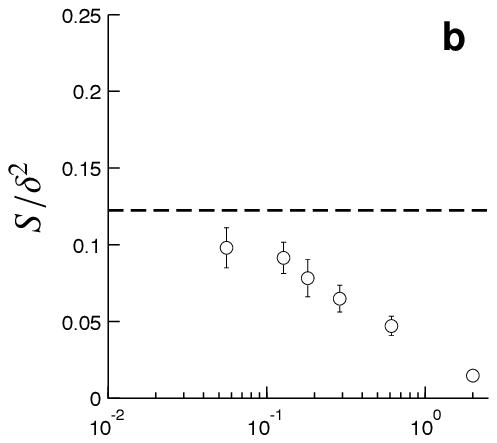}\\[0.7em]
   \includegraphics[width=0.35\columnwidth,height=0.31\columnwidth]{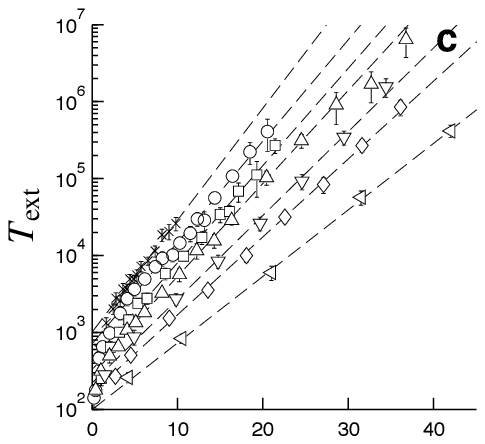}\qquad
   \includegraphics[width=0.35\columnwidth,height=0.31\columnwidth]{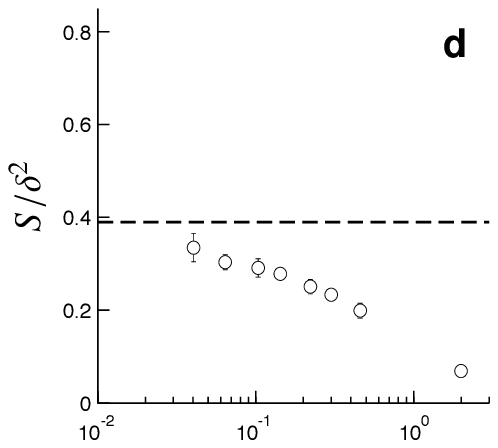}\\[0.7em]
   \includegraphics[width=0.35\columnwidth,height=0.35\columnwidth]{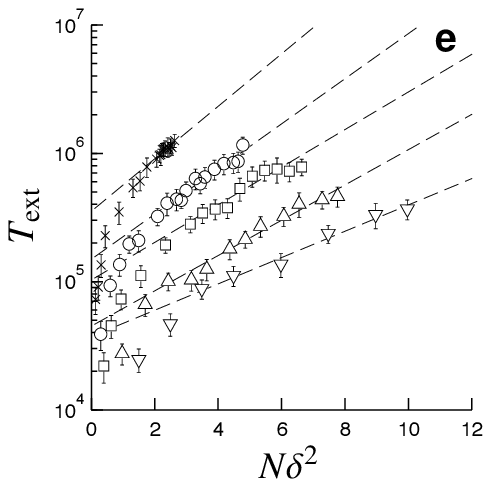}\qquad
   \includegraphics[width=0.35\columnwidth,height=0.35\columnwidth]{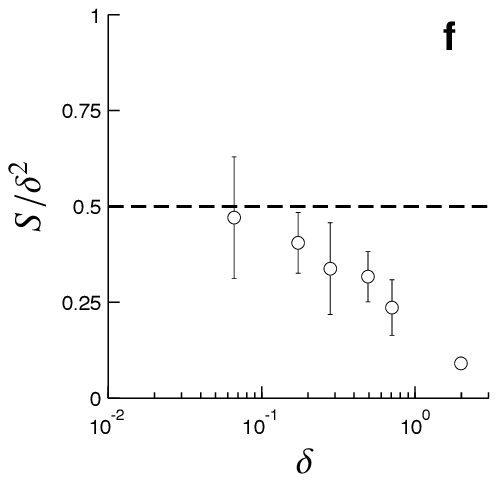}
   \caption{\label{fig:Text} (\textsf{\textbf{a}}) Average time to extinction $T_{\rm ext}$ from direct numerical simulations of the model described 
    in section \ref{sec:model}, as a function of the number of patches $N$, for $r=1.05$, $K=10$. Symbols 
    ($\times$, $\bigcirc$, $\square$, $\triangle$, $\triangledown$) 
     correspond to the following values of $\delta$: $0.0557$, $0.128$, $0.182$, $0.289$, $0.611$. Each data point corresponds to an average over 100 stochastic realisations.
     Also shown are numerical fits to the expected behaviour Eq.~(\ref{eq:expNS}).
    (\textsf{\textbf{b}}) Action as a function of $\delta$. Shown are the limiting result, Eq.~\eqref{eq:action2} (dashed line), 
    as well as the results of the fits shown in the left panel ($\bigcirc$). Error bars correspond to 95\% confidence intervals.
    (\textsf{\textbf{c}}-\textsf{\textbf{d}}) Same as top panels, but for $r=1.5$ and $K=10$.
    The symbols ($\times$, $\bigcirc$, $\square$, $\triangle$, $\triangledown$, $\Diamond$, $\vartriangleleft$) 
    correspond to $\delta=0.04$, $0.064$, $0.103$, $0.143$, $0.222$, $0.301$, and $0.458$.
    (\textsf{\textbf{e}}-\textsf{\textbf{f}}) Same as top panels, but for $r=1.5$ and $K=50$.
    The symbols ($\times$, $\bigcirc$, $\square$, $\triangle$, $\triangledown$) correspond to $\delta=0.066$, $0.173$, $0.279$, $0.493$, and $0.706$.  
    }
\end{figure}

%----------------------------------------------------------------------------------------------------------------
The tail of the quasi-steady state distribution (\ref{eq:ansatz}) towards $\ve f =\ven 0$ determines
the average time to extinction, Eq.~(\ref{eq:Text0}):
\begin{equation}
\label{eq:expNS}
T_{\rm ext} = A\, {\rm exp}\big(N S\big)\,.
\end{equation} 
The coefficient $A$ may depend on $N$, as well as on $r, K$, and $m$. 
We have not been able to determine it.
The action $S\equiv S(\ve f = \ven 0)$ is a function of $r, K$, and $m$.
Since $N S$ appears in the argument of the exponential
in Eq.~(\ref{eq:expNS}), the average time to extinction depends sensitively
upon the number $N$ of patches, and on $S$. 
Close to the critical line in Fig.~\ref{fig:mc} we find the action by integrating
$P_1$ along the path given by Eq.~(\ref{eq:instanton}). This yields:
\begin{equation}
\label{eq:action2}
S =-\frac{\delta^2}{2} \frac{ a_{11}^2}{b_{11}a_{12}}
\end{equation}
which corresponds to the shaded area in Fig.~\ref{fig:PQ_plane}.
Note that $a_{12} < 0$.
The prediction for $S$, Eq.~(\ref{eq:action2}), is compared to results of direct simulations in Fig. \ref{fig:Text}.
The initial condition for the direct simulations was chosen as follows. First $\ve f^*$ was calculated,
this determines  the initial state $\ve n = N \ve f^*$. Note however that the components $n_j$ of $\ve n$ must be integers. 
For small values of $\delta$, all $n_{j>0}$ may round to zero, inconsistent with the constraint $\sum_j n_j = N$.
In such cases we grouped counts $n_j$ corresponding to neighbouring values of $j$ together before rounding. 
The action $S$ was determined by plotting $\log T_{\rm ext}$ as a function of $N$. For sufficiently large values of $N$
(such that $N S\gg 1$) one expects a straight line. We perform a linear regression by least squares to determine the slope $S / \delta^2$. 
Fig.~\ref{fig:Text} shows $\log T_{\rm ext}$ as a function of $N\delta^2$.
The approximations leading to Eq.~(\ref{eq:action2}) require that $N$ is large and $\delta$ small. Eq.~(\ref{eq:expNS}) is expected
to be a good approximation provided
\begin{equation}
(NS)^{-1/2} \ll \delta \ll 1\,.
\end{equation}
The data shown in Fig.~\ref{fig:Text} are consistent with this expectation.
We see that the fitted values of $S$ approach the analytical result, Eq.~(\ref{eq:action2})
as $\delta$ is decreased (right panels in Fig.~\ref{fig:Text}).  This figure
illustrates the sensitive dependence of the time to extinction
upon $r$, $K$, $\delta$, and $N$.

\begin{figure}[htp]
   \centering
   \includegraphics[width=0.4\columnwidth,height=0.4\columnwidth]{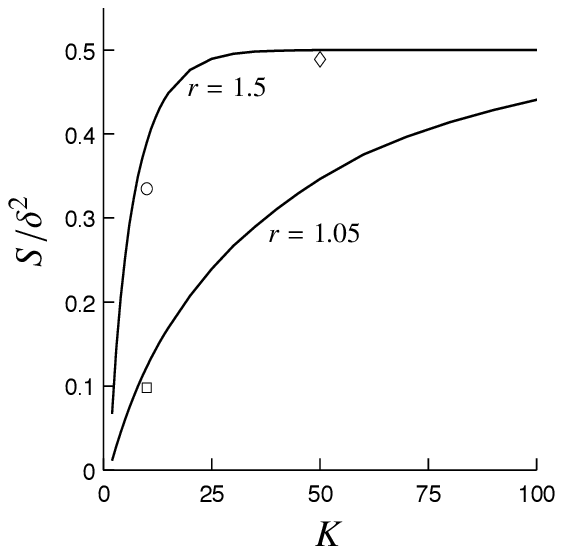}\qquad
      \caption{\label{fig:action} Action $S$ divided by $\delta^2$ according to Eq.~(\ref{eq:action2}), in the limit of $\delta \rightarrow 0$, as a function of the carrying capacity $K$. Solid lines correspond to $r = 1.05$ and $1.5$. 
Symbols ($\square$, $\bigcirc$, $\Diamond$)
correspond to the best estimates from Fig.~\ref{fig:Text}a, Fig.~\ref{fig:Text}c, and Fig.~\ref{fig:Text}e.}
\end{figure}
In Fig.~\ref{fig:action}  we summarise our best estimates for $S/\delta^2$ for different values of $r$ and $K$ 
(and for the smallest value of $\delta$ for which we obtained reliable results). These estimates
are compared to the analytical result, Eq.~(\ref{eq:action2}). We observe good agreement.
Fig.~\ref{fig:action} appears to indicate that $S/\delta^2$ approaches the value $1/2$ as $K$ increases.
The approach appears to be the faster the large the value of $r$ is. 
In order to demonstrate that this is in fact true, we have determined the
asymptotic dependency of the coefficient $b_{11}$ upon
$K$ in the limit of large values of $K$. We find
(see appendix \ref{app:C}):
\begin{equation}
\label{eq:asy}
b_{11} \sim  \sqrt{\frac{K(r-1)^3}{2\pi }}
\exp\Big[-K\Big(1-\frac{\log r}{r-1}\Big)\Big]\,.
\end{equation}
This result, taken together with Eq.~\eqref{eq:a11_a12_asy} determines the limiting value of $S/\delta^2$ to be $1/2$.
Fig.~\ref{fig:action} shows that already for $r=1.5$ and $K=25$ the value of $S/\delta^2$ is very close
to this limiting value.  We note that while the expressions (\ref{eq:a11_a12_asy}) and (\ref{eq:asy}) depend upon $K$, the action (\ref{eq:action2}) approaches
a limit independent of $K$ for large values of $K$. In this limit, the metapopulation dynamics can be understood
in terms of a stochastic dynamics of the fraction of occupied patches. 

Such an approach was suggested by \citet{Lande1998}. In the following we briefly describe
the similarities and differences between our asymptotic result and the approach suggested by \citet{Lande1998}.
As already pointed out in the introduction, their analysis rests on four main assumptions:
First, \citet{Lande1998} consider the limit of fast local dynamics
and slow migration (time-scale separation). In our model, this corresponds
to the limit $K\rightarrow \infty$ and to large values of $r$.
Second, \citet{Lande1998} argue that the extinction rate in the stochastic model for
the evolution of the number of occupied patches is given by the inverse time to extinction of a single patch, initially at carrying capacity. 
Third, it is assumed that the rate of successful colonisation is given by the product of the expected
number of migrants and the probability that an empty patch invaded by one migrant grows
to its (quasi-)steady state. Fourth, \citet{Lande1998} allow for the possibility
that the number of individuals per patch in occupied patches changes as
the metapopulation approaches extinction, and argue that this effect can
be incorporated in terms of $Q$-dependent rates $c(Q)$ and $e(Q)$.  

Here we have shown, however, that in the limit of small values of $\delta$ the dynamics of $\ve f$ on the most
likely path to extinction is rigid: $\ve f$ does not change its shape, just
its normalisation. This implies, in particular, that the average number of individuals
on occupied patches must remain unchanged as extinction is approached.
This is clearly seen in Fig.~\ref {fig:colorprob}a. 
We note, however, that Fig. \ref {fig:colorprob}b shows that for larger values of $\delta$, 
$\ve f$ does change its shape (and the average number of individuals in occupied patches
decreases as extinction is approached). A first-principles theory for this
effect is lacking. We refer to this point in the conclusions.

Let us now consider the parameterisations of the rates $e$ and $c$
suggested by \citet{Lande1998}, Eqs.~(3) and (4) in their paper.
We have obtained analytical results
in the vicinity of the bifurcation (that is, for small values of $\delta$).
In order to compare our results to the choices adopted by \citet{Lande1998} we must
take the limit $K\rightarrow \infty$. In this limit, it turns out, that the terms
of order $\delta^3$ (and higher) in the second and third rows of Eq. (\ref{eq:Le0})
are negligible compared to the terms in the first row. In the limit
of large values of $K$ we therefore conclude:
\begin{equation}
c = -(1+\delta) a_{12}\quad \mbox{and } \quad e = c  -a_{11}\delta\,.
\end{equation}
The asymptotic expressions for $a_{11}$ and $a_{12}$ are given in Eq.~(\ref{eq:a11_a12_asy}).
In appendix \ref{app:C} we have shown
that these relations correspond precisely 
to the asymptotic $K$-dependence of the time $T_K$ to extinction
for a single isolated patch at carrying capacity. In the limit of large
$K$ (which is the limit \citet{Lande1998} consider) we thus have:
\begin{equation}
\label{eq:eTK}
e = \frac{1}{T_K}\,.
\end{equation}
This equation implies that the rate of extinction of patches
in the coupled system does not depend upon $\delta$ to leading order in $\delta$.
In other words, patches go extinct independently in the limit $K\rightarrow \infty$.
Eq.~(\ref{eq:eTK}) closely resembles Eq.~(4) in \citep{Lande1998}.

Now consider the rate of successful colonisation. In the limit of $K\rightarrow\infty$,
the probability of a single patch growing from one migrant to carrying capacity
is $u_{1K} \sim (r-1)/r$ (see appendix \ref{app:C}). Immigration is irrelevant in this context since the local
patch dynamics is assumed to be much faster than migration. Using
this expression for $u_{1K}$ and (\ref{eq:mcexpand}) for $m_{\rm c}$ we find the following expression for $c$
\begin{equation}\label{eq:cLande}
c = m_{\rm c} (1+\delta) K u_{1K}=m K u_{1K}\,.
\end{equation}
This result closely resembles Eq.~(3) in \citep{Lande1998}.
We emphasise that our results are exact in the limit of small values of $\delta$
and large values of $N$ and $K$. 

A further but minor difference
is that \citet{Lande1998} employ the diffusion approximation to
evaluate their Eqs.~(3) and (4). This approximation fails unless $r$ is close to unity.

We conclude this section with a discussion of Eq.~(\ref{eq:action2}).  Eqs.~(\ref{eq:fR},\ref{eq:pR}) 
allow us to express the average time to extinction in the following form:
\begin{equation}
\label{eq:Text2}
\log T_{\rm ext} \sim  \frac{N}{2} (-\ve p^{\ast \sf T}) \ve f^\ast
=  \frac{N}{2} \sum_{j=1}^\infty (-p_j^\ast) f_j^\ast\,.
\end{equation}
Here $\ve f^\ast$ is the quasi-steady state distribution, and the
vector $-\ve p^{\ast \sf T}$ determines the susceptibility of patches
with $j$ individuals to stochastic fluctuations of the number of individuals.
As explained above, the components $-p_j^\ast$ 
are related to Fisher's reproductive
vector characterising the susceptibility, for example, of age classes
in matrix population-models. 
But note that in our case the components of $-\ve p^{\ast}$ 
do not characterise the properties of classes of individuals, but of patches.
In other words, Eq.~\eqref{eq:Text2} can be understood by viewing the metapopulation
as a population of patches, or as a population of local populations, 
as originally envisaged by \citet{Levins1969}.

%*******************************************************************************************
\subsubsection{The limit of large emigration rates}
%*******************************************************************************************
\label{app:D}
Most results presented in this paper concern metapopulation dynamics 
close to the critical line shown in the left panel of Fig.~\ref{fig:mc}. The reason is that
the metapopulation dynamics simplifies substantially in this regime, as shown in
the preceding sections.

In this section we briefly discuss another limit of the metapopulation dynamics in 
the model introduced in Section~2, the limit of large emigration rates.
This limit corresponds to the top part of the $m$-$K$ plane  shown in 
the left-most panel of Fig.~\ref{fig:mc}.
The limit is interesting for two reasons. First, in this limit too the metapopulation
dynamics simplifies substantially. We show that the dynamics can be represented
in terms of a one-dimensional process (just as Eq.~(\ref{eq:LevinsModel}) and
the corresponding stochastic dynamics). But now the variable is the total
number of individuals in the metapopulation and not $Q_1$, or $Q$.
Second, the results allow us to connect the subject and the results of the present paper to
recent results by \cite{higgins_metapopulation_2009}.

In the limit of infinite emigration rate ($m\rightarrow \infty$) and
no mortality during migration ($\zeta=0$), the metapopulation
behaves as a single large patch with modified birth and death rates that depend only on the total number of individuals in the metapopulation.
This makes it possible to compute the average time to extinction exactly.
This limit could describe, for example, biological populations in which larvae disperse randomly to all patches (e.g. pelagic marine species).

We start from the individual-based, stochastic metapopulation model introduced in Section~2, 
in the limit $\eta\rightarrow\infty$ and for $\zeta=0$.
There are $N$ patches.  It is convenient to characterise the state of the metapopulation by the number of individuals in each patch $i_1,\ldots,i_N$
(rather than by the numbers $f_j$ denoting the fraction of patches with $j$ individuals as in the preceding sections).
Since migration is infinitely faster than any other process, migration instantly brings the distribution of $i_1,\ldots, i_N$
to multinomial form. 
Denoting the total number of individuals in the metapopulation by $n$,
we see that after a birth ($n-1\rightarrow n$) or death ($n+1\rightarrow n$), the distribution
instantaneously relaxes to the multinomial distribution:
\begin{equation}
\label{eq:mn}
         M_n(i_1,\ldots,i_N)=
\left\{\begin{array}{ll}
\frac{n!}{i_1! i_2! \ldots i_N} N^{-i_1}  N^{-i_2} \cdots  N^{-i_N} & \mbox{if $\sum_{k=1}^N i_k = n$\,,}\\[0.1cm]
0 & \mbox{otherwise.}
\end{array} \right .
\end{equation}
We denote the birth and death rates for the whole metapopulation, conditional upon $n$, by  $B_n$ and $D_n$, respectively.
These rates can be calculated by averaging the local birth and death rates, Eqs.~(\ref{eq:rates}) and (\ref{eq:dr}), 
conditional upon $n$. For the birth rate we find
\begin{equation}\label{eq:Bn}
B_n = \Big\langle \sum_{k=1}^N b_{i_k} \Big\rangle = r \,\Big\langle \sum_{k=1}^N i_k\Big\rangle = rn\,,
\end{equation}
by virtue of $\sum_{k=1}^N i_k = n$. For the death rate we find:
\begin{align}\label{eq:Dn}
        D_n = {}&\!\! \sum_{i_1,\ldots,i_N} \Big (\sum_{k=1}^N d_{i_k}\Big)\, M_n(i_1,\ldots,i_N) = \mu n + \frac{r\!-\!\mu}{NK}n^2 + \frac{r\!-\!\mu}{K}n \left(1\!-\!\frac{1}{N}\right)\,.
\end{align}
These rates (we take $\mu=1$ as before) define a one-dimensional, one-step
birth-death process, that can be written in terms of a one-dimensional one-step  master equation
\begin{equation}\label{eq:masterminf}
	\frac{\dd \rho_n}{\dd t} = (\mathbb{E}^- -1)B_n\,\rho_n + (\mathbb{E}^+ -1)D_n\,\rho_n\, ,
\end{equation}
for the probability $\rho_n(t)$ of finding $n$ individuals in the metapopulation at time $t$. 
The raising and lowering operators in Eq.~(\ref{eq:masterminf}) are defined in analogy with Eq.~(\ref{eq:rl}):
$\mathbb{E}^{\pm} g_n = g_{n \pm 1}$. The average time to extinction
for the one-dimensional Eq.~(\ref{eq:masterminf}) can be obtained by recursion (the result is given in appendix~C).
For the case of the multi-dimensional master equation (\ref{eq:mastereq}), by contrast, it was
necessary to employ a large-$N$ expansion before the dynamics could be reduced
to the one-dimensional form (\ref{eq:LevinsModel}) on the brink of extinction.

\begin{figure}[htp]
   \centering
   \includegraphics[width=0.4\columnwidth,height=0.4\columnwidth]{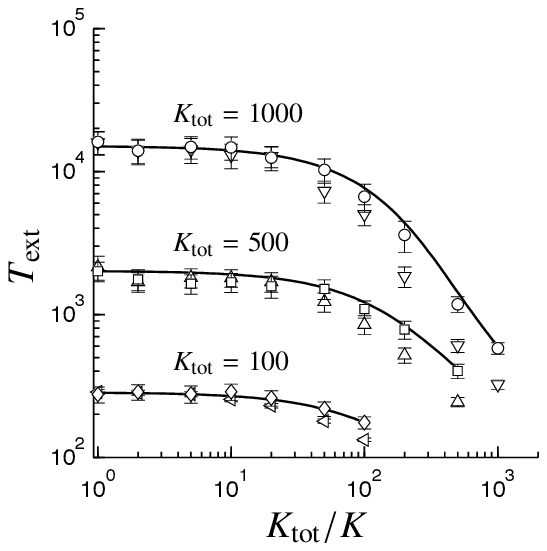}
\caption{\label{fig:Textinfm}
 Average time to extinction, $T_\textrm{ext}$, as a function of the degree of fragmentation $K_{\rm tot}/K$. 
 See Section \ref{app:D}.
 Parameters:  $r = 1.01$, $m=1,10$, $K_{\rm tot} = 100,500,1000$.
Shown are results of Eqs.~(\ref{eq:Texact},\ref{eq:Bn},\ref{eq:Dn}) (solid lines) and results of  direct numerical simulations as described in Sec.~\ref{app:D} for $m=10$ 
($\bigcirc$, $\square$, $\Diamond$) and $m=1$ ($\triangledown$, $\triangle$, $\vartriangleleft$). Error bars correspond to 95\% confidence intervals.
 }
\end{figure}

The results (\ref{eq:Bn}-\ref{eq:masterminf}) allow us to investigate the question addressed by \cite{higgins_metapopulation_2009},
namely how the risk of extinction depends upon the degree of \lq fragmentation' of a metapopulation,
in the limit of infinite emigration rate.
We define the total carrying capacity of the metapopulation $K_{\rm tot}  = KN$ and ask
how the average time of extinction depends upon $N$, keeping $K_{\rm tot}$ constant.
In other words, we fragment the metapopulation into $N$ patches of equal size, such that the total carrying capacity remains the same.  Figure~\ref{fig:Textinfm} shows how the 
average time to extinction depends on the degree of fragmentation, for three different values of $K_{\rm tot}$ 
(100, 500, and 1000), for $r = 1.01$ (other parameter values give qualitatively similar results).
The initial number of individuals is taken to be $5 K_{\rm tot}$ (the average time to extinction depends only weakly
on the initial number of individuals, provided this number is larger than $K_{\rm tot}$). 
The analytical results (lines) are in good agreement with results, for large emigration rates, 
of direct numerical simulations of the model introduced in Section~2. Results are shown for 
two values of the emigration rate, $m=1$ and $10$.  Each data point (symbols in Fig.~\ref{fig:Textinfm}) 
is estimated from $100$ independent simulations.

We find that the time to extinction decreases monotonically as the degree of fragmentation increases.
For small degrees of fragmentation, the time to extinction is approximately independent of the number of patches, but for larger levels 
of fragmentation, the time to extinction decreases markedly as fragmentation increases. 
Since the time to extinction is approximately exponentially distributed, this implies that the probability 
of extinction during a given time span increases monotonically as the metapopulation becomes more and more fragmented. 

This result is in qualitative agreement with the findings of \citet{higgins_metapopulation_2009} for intermediate and large degrees
of fragmentation.  By contrast, when the number of patches is small, \citet{higgins_metapopulation_2009} 
observes a decrease in the extinction risk with increasing fragmentation.
A possible reason for this difference is that \citet{higgins_metapopulation_2009} considers local environmental fluctuations that would increase the local rate of extinction compared to demographic fluctuations alone \citep{MelbourneHastings:2008}. 
This could be important when the number of patches is low, but when the number of patches is large the fluctuations are averaged over 
and may become less important.

%****************************************************************************************************************
\section{Conclusions}
\label{sec:c}
%****************************************************************************************************************

In this paper we have analysed metapopulation dynamics on the brink of extinction
in terms of a stochastic, individual-based metapopulation model consisting
of a finite number $N$ of patches. In our model the distance from the bifurcation point
where infinitely large metapopulations cease to persist is parameterised
by the parameter $\delta$. It measures the difference in emigration rate to its critical value at the bifurcation point. In the limit of large (but finite) values of $N$
and small values of $\delta$ we have been able to 
quantitatively describe the stochastic metapopulation dynamics.
We have shown that metapopulation dynamics, for small values of $\delta$, is
described by a one-dimensional equation of the form of Levins' model, Eq.~(\ref{eq:LevinsModel}).
We have derived explicit expressions for
the parameters $c$ and $e$ in terms of the parameters
describing the local population dynamics, and migration. We have shown 
that Levins' model is valid independently of whether or not
there is a time-scale separation between local and migration dynamics.
The crucial condition is that the metapopulation is close to extinction. 
We note that in the absence of time-scale separation, the interpretation
of the variable $Q$ in Eq.~(\ref{eq:LevinsModel}) is no longer the fraction of occupied patches.
More precisely, $Q$ corresponds to the fraction of occupied patches if the growth rates and local carrying capacities are large enough; otherwise it has a different meaning, in particular, for small growth rates and carrying capacities it approximately corresponds to the average number of individuals per patch.

\begin{figure}[htp]
   \centering
    \includegraphics[width=0.98\columnwidth]{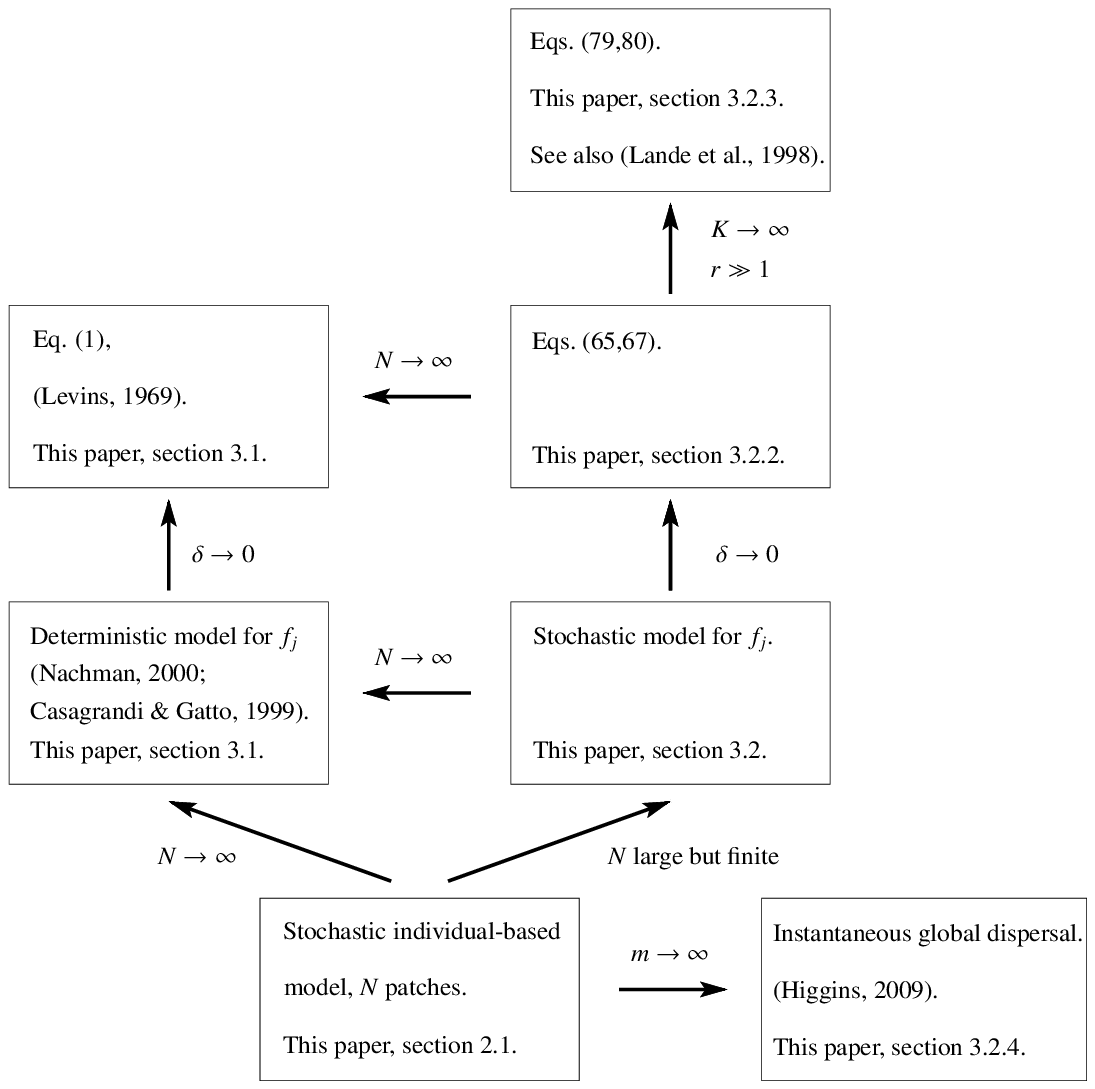}
      \caption{\label{fig:15} Schematic diagram showing relations between different limiting cases of the individual-based stochastic metapopulation model
    described in section \ref{sec:model}. The metapopulation consists of $N$ patches coupled by migration.
    The variables $f_j$ denote the fraction of patches containing $j$ individuals.
     The parameter $m$ denotes the emigration rate, Eq.~(\ref{eq:rates2}),
    and $\delta$ characterises how $m$ differs from the critical rate $m_{\rm c}$,
    see Eqs.~(\ref{eq:defdelta},\ref{eq:mc})  in Sec.~\ref{sec:detdyn}.
   }
\end{figure}

The deterministic limit of our model corresponds to metapopulation models discussed by \citet{Casagrandi2002} and \citet{Nachman:2000}, who have
studied the persistence of infinitely large metapopulations by analysing
the stability of steady states. The corresponding persistence criteria
(which we discuss in detail) do not allow to characterise the persistence
of finite metapopulations. 
By contrast, the stochastic dynamics derived
and analysed in this paper makes it possible to characterise
the most likely path to extinction and to estimate the
average time to extinction of the metapopulation. 
We have discussed differences and similarities between
a particular asymptotic limit of our results, and
the results obtained by \citet{Lande1998}.
Fig.~\ref{fig:15} summarises different limiting cases of
our results, and connections to earlier studies.

While our approach results in a comprehensive description
of metapopulation dynamics for the model we consider,
many open questions remain. A technical point is
that we have not yet been able to compute
the pre-factor $A$ in the expression (\ref{eq:expNS}) 
for the average time to extinction. This is a difficult
problem requiring matching the solutions found
in this paper with corresponding solutions
valid for small values $n_j$ (the number of
patches with $j$ individuals). Such solutions
are outside the scope of our large-$N$ treatment.

Many results derived in this paper
are valid close to the critical line in Fig.~\ref{fig:mc}, that is on the
brink of extinction. The reason is that the metapopulation dynamics simplifies
considerably in this regime, as we have shown in this paper. Further work is required in order to understand
metapopulation dynamics for emigration rates much larger than  the critical value. In this
case, especially when the number of migrants is large, we have observed deviations from Levins' model:  
the shape of the distribution of local patch population sizes is no longer rigid on the path to extinction. Biologically this reflects, for example, the rescue effect, whereby immigration from colonised patches extends the time to local extinction.
A quantitative theory of this effect based on the life history of the individuals in the metapopulations is currently lacking.

There is, however, a second limit of our individual-based, stochastic metapopulation model 
that can be analysed in a simple fashion. This is the limit of very large emigration rates, $m\rightarrow \infty$,
where the metapopulation is well mixed, so that, in the absence of mortality during migration, it can be modelled as a single large patch. In this limit
it is possible to analyse the question posed by \citet{higgins_metapopulation_2009}, namely how the risk of extinction
of a metapopulation depends on the degree of its fragmentation.
In the limit of large emigration rates, we have derived the expected time to extinction
for our individual-based stochastic metapopulation model. 
We find that the time to extinction decreases as the degree of fragmentation of the population increases.
There is excellent agreement between this prediction and results
of direct numerical simulations of the model introduced in Section~2, for large but finite emigration rates.
For highly fragmented populations, our conclusions are consistent
with the results of \citet{higgins_metapopulation_2009}. 
When the number of patches is small, by contrast, \citet{higgins_metapopulation_2009} observes a 
different behaviour (discussed in Sec.~\ref{app:D}).
A possible reason for this difference is that \citet{higgins_metapopulation_2009} considers local environmental fluctuations. These would increase the local rate of extinction compared to demographic fluctuations alone \citep{MelbourneHastings:2008}.
This could be important when the number of patches is low. 
But when the number of patches is large the fluctuations are averaged over and are thus expected to be less important.
To understand the differences and similarities between the models (and their biological implications) requires further work.

The approach described in this paper allows us to address many 
other important questions in metapopulation dynamics. Consider for example the dynamics of structured metapopulations. 
Core-satellite models \citep{hanski_gyllenberg93} of metapopulations have two types of patches, large and small.
Large patches have more persistent populations, while populations in smaller patches are relatively ephemeral.
Thus, the large patches are, on average, sources and the small patches are sinks for the metapopulation as a whole.
Such models can be treated by suitable generalisations of the approach described in the present paper.

Another important question is the role of environmental fluctuations. 
The within-patch population dynamics in our model takes demographic
fluctuations into account, but does not explicitly incorporate the effect of local
environmental fluctuations. Such fluctuations may be due to,
for example, yearly fluctuations in growth rates, mortality, or
carrying capacity \citep{MelbourneHastings:2008}. The environmental fluctuations
may be local, affecting patches independently. Alternatively the fluctuations
of the rates may be global, that is, the same for all patches (or at least
highly correlated).

What is the effect of local environmental fluctuations? 
Consider the set of patches with $j$ individuals. 
One may expect that, when the number of patches is large, only 
average rates enter into the deterministic equations of motion (\ref{eq:det}).  
But the environmental noise gives rise to fluctuations, and therefore
leads to shorter persistence times of individual patches. This in turn
implies higher patch turnover \citep{MelbourneHastings:2008}. 

When the patches are few, the destabilising effect of local environmental fluctuations on patches can
significantly affect the time to extinction \citep{higgins_metapopulation_2009}.
The tendency of local environmental fluctuations to increase patch turnover
could be represented in our model by introducing a separate process to the
local population dynamics, whereby patches go extinct at a fixed rate,
independently of the local population size, for example in terms of 
the `killing' process \citep{coolen_schrijner_2006}. 
The theory outlined in this paper then makes it possible to determine
the effect of this process upon the time to extinction of the metapopulations.

Global environmental fluctuations, affecting all patches simultaneously, can be incorporated in our model by introducing a time-dependent contribution $\delta \mu(t)$ to the death rate in Eq. (3), the same for all patches. Taking this to be a random piecewise-constant function of time makes it possible to employ the approach described by \citet{Raf}.

Last but not least, it is necessary to compare the predictions summarised
here with those of models with an explicit spatial structure. We expect
that the predictions summarised here should be the more accurate the
wider the spatial scale of migration is.

{\em Acknowledgements}.
Financial support by Vetenskapsr\aa{}det, by the Centre for Theoretical Biology at the University of Gothenburg, 
and by the G\"oran Gustafsson Foundation for Research in Natural Sciences and Medicine are gratefully acknowledged.

%%\newpage
%\onecolumn
\begin{appendices}
\numberwithin{equation}{section}
%****************************************************************************************************************
\section{Formulae for the matrix elements occurring in the expansion of the master equation}
%****************************************************************************************************************
\label{app:A}
The matrix ${\ma D}(\ve f)$ (see Eq.~\eqref{eq:H2}) has entries
\begin{alignat}{2}
\label{eq:partial_D}
        D_{ij} = {} & \delta_{ij}\Big( (b_{i-1}+I) f_{i-1} + (d_{i+1}+m_{i+1}) f_{i+1} + (b_i+I+d_i+m_i)f_i \Big)   \\
        &+ \delta_{i+1 j}\Big( -(d_{i+1}+m_{i+1}) f_{i+1}  - (b_i+I)f_i \Big) \nn \\
        &+ \delta_{i-1 j}\Big( -(b_{i-1}+I) f_{i-1} - (d_i+m_i)f_i   \Big)  \nn \\
        &+ m_{i+1}f_{i+1}(f_{j-1} - f_j) + f_i(m_jf_j - m_{j+1}f_{j+1}) \nn \\
        &+ m_{i}f_{i} (f_j-f_{j-1}  ) + f_{i-1} (m_{j+1}f_{j+1}-m_jf_j ) && \hspace{-0.7cm}{\mbox{for $i,j > 1$}}\, , \nn \\[1em]
        D_{1j} = {} & \delta_{j2}\Big( -(d_{2}+m_{2}) f_{2}  - (b_1+I)f_1 \Big) + m_{2}f_{2}(f_{j-1} - f_j) +  m_{1}f_{1} (f_j-f_{j-1}  )   \nn\\
        & + f_1(m_jf_j - m_{j+1}f_{j+1}) +   \left(1-\sum_{k=1}^\infty f_k\right)(m_{j+1}f_{j+1}-m_jf_j ) &&\hspace{-0.7cm}{ \mbox{for $j > 1$}}\, , \nn \\
        D_{i1} = {} & \delta_{i2}\Big( -(d_{2}+m_{2}) f_{2}  - (b_1+I)f_1 \Big)  + m_{2}f_{2}(f_{i-1} - f_i) +  m_{1}f_{1} (f_i-f_{i-1}  )  \nn\\
        &+ f_1(m_if_i - m_{i+1}f_{i+1}) +   \left(1-\sum_{k=1}^\infty f_k\right)(m_{i+1}f_{i+1}-m_if_i ) && \hspace{-0.7cm}{\mbox{for $i > 1$}}\, , \nn \\
        D_{11} = {} & I \left(1-\sum_{k=1}^\infty f_k\right)  + (d_{2}+m_{2}) f_{2} + (b_1+I+d_1+m_1)f_1 \nn \\
        &+ 2\left(1-\sum_{k=1}^\infty f_k - f_1\right)(m_2f_2 - m_1f_1)\,.\nn
\end{alignat}
The elements of ${\ma A}$ (see Eq.~\eqref{eq:J}) are given by
\begin{alignat}{2}
\label{eq:A}
A_{ij} = {}& (b_{i-1}+I) \delta_{i-1j} + (d_{i+1}+m_{i+1}) \delta_{i+1j} 
-(b_i+I+d_i+m_i)\delta_{ij} \\
&+ m_j(f_{i-1} - f_i)  && \mbox{for $i> 1$}\,, \nn \\
A_{1j}= {}&  (d_{2}+m_2) \delta_{j2} -(b_1+I+d_1+m_1)\delta_{j1} - I + m_j\Big(1-\sum_{k=1}^{\infty}f_k - f_1\Big)\, . \quad && \mbox{} \nn
\end{alignat}
The elements of ${\ma A}^{(0)}$ in the expansion \eqref{eq:expand} are obtained by setting $\delta=0$ in Eq.~(\ref{eq:A}).
The elements of ${\ma A}^{(1)}$ occurring in the same expansion are found to be:
\begin{equation}\label{eq:partial_v_dfdm}
A_{ij}^{(1)}  = \left\{ \begin{array}{ll}
j( \delta_{i+1j}-\delta_{ij}) \quad
& \mbox{for $i>1$}\\
j(\delta_{j2} - \delta_{j1} + 1) & \mbox{for $i=1$}
\end{array}
\right . \,.
\end{equation}
Finally, the coefficients ${A}^{(2)}_{ijk}$, also appearing in the expansion \eqref{eq:expand},  are given by:
\begin{equation}
\label{eq:partial_v_df2}
A_{ijk}^{(2)}  =
\left\{ \begin{array}{ll}
- k (\delta_{ij}-\delta_{i-1j})
- j (\delta_{ik}-\delta_{i-1k}) \quad
& \mbox{for $i>1$}\\
- k(\delta_{1j}+1)-j(\delta_{1k}+1) & \mbox{for $i=1$}
\end{array}
\right . \,.
\end{equation}
The coefficients ${B}^{(1)}_{ijk}$ (see Eq.~\eqref{eq:Bijk}) are given by:
\begin{alignat}{2}
        B_{ijk}^{(1)} = {} & b_{k}\big[ \delta_{ki-1}(\delta_{ij} - \delta_{ji-1} ) +\delta_{ik}(\delta_{ij} - \delta_{ji+1} ) \big]   \\
                                                        & + (d_{k} + m_{k}) \big[ \delta_{ki+1}(\delta_{ij} - \delta_{ji+1} ) + \delta_{ik}(\delta_{ij} - \delta_{ji-1} ) \big] \qquad&& \mbox{for $i,j>1$}, \nn\\
        B_{1jk}^{(1)} = {} & (d_2 + m_2)(-\delta_{k2}\delta_{j2}) - b_1\delta_{k1}\delta_{j2} + m_{k}(\delta_{kj+1} - \delta_{kj}) && \mbox{for $j>1$}, \nn\\
        B_{i1k}^{(1)}  = {} & (d_2 + m_2)(-\delta_{k2}\delta_{i2}) - b_1\delta_{k1}\delta_{i2} + m_{k}(\delta_{ki+1} - \delta_{ki}) && \mbox{for $i>1$}, \nn\\
        B_{11k}^{(1)}  = {} & m_k + (d_2 + 3m_2)\delta_{k2} + (b_1 + d_1 - m_1)\delta_{k1} .\nn
\end{alignat}

%****************************************************************************************************************
\section{Summary of results for coefficients used in section~\ref{sec:results}}
%****************************************************************************************************************
\label{app:B}
In this appendix we give explicit expressions for the coefficients $I_1$, $c_1$, $c_2$, $a_{11}$, $a_{12}$ and $b_{12}$
appearing in Eqs.~\eqref{eq:Iexpand}, \eqref{eq:f0}, \eqref{eq:mu}, \eqref{eq:Le1}, and \eqref{eq:slowH}. 
First, the coefficient $I_1$ in Eq. (\ref{eq:Iexpand}) is calculated by expanding Eq.~\eqref{eq:sc}. 
Expanding to lowest order yields the condition (\ref{eq:mc}) which gives $m_{\rm c}$ but does not determine $I_1$.
In order to find  $I_1$ it is necessary to expand Eq.~(\ref{eq:sc}) to second order in $\delta$. This requires
expanding:
\begin{align}
\label{eq:S1}
m_{\rm c}(1+\delta) \sum_{j=2}^\infty j\prod_{k=2}^j\frac{b_{k-1}+I^\ast}{d_k+m_{\rm c} k}  
= {} & I_1 \delta -I_2 \delta^2 + g_1(I_1) \delta^2+\ldots\,,\\
\sum_{j=2}^{\infty}
\prod_{k=2}^j\frac{b_{k-1}+I^\ast}{d_k+m_{\rm c} k} 
={}& g_0(I_1) \delta+\ldots \,.\nn
\end{align}
We find:
\begin{align}
g_0(I_1) = {}& \frac{I_1}{d_1 + m_{\rm c}} \Big(1+\sum_{j=2}^\infty \frac{R_{1j}}{R_{11}}\Big)\\
g_1(I_1) = {}& \frac{d_1}{d_1+m_{\rm c}} I_1 -\frac{m_{\rm c}I_1}{d_1+m_{\rm c}} \sum_{k=2}^\infty  \frac{kR_{1k}}{R_{11}} 
\sum_{i=2}^k \frac{m_{\rm c}i}{d_i+m_{\rm c}i} \\
&+ \frac{m_{\rm c} I_1^2}{d_1+m_{\rm c}}\sum_{k=2}^\infty \frac{kR_{1k}}{R_{11}}  \sum_{i=2}^k \frac{1}{b_{i-1}}\,. \nn
\end{align}
Inserting these expansions into the self-consistency condition (\ref{eq:sc}) gives:
\begin{equation}
I_1 = \frac{g_1(I_1)}{g_0(I_1)}\,.
\end{equation}
We find
\begin{equation}
I_1 = \frac{d_1 -m_{\rm c} \sum_{k=2} \frac{kR_{1k}}{R_{11}}  \sum_{i=2}^k \frac{m_{\rm c}i}{d_i + m_{\rm c}i}}{1+\sum_{k=2}^\infty \frac{R_{1k}}{R_{11}} (1-k\sum_{i=2}^k \frac{1}{b_{i-1}})}\,.
\end{equation}
Second, the coefficient $c_1$ determines the leading order of the  $\delta$-expansion of the fraction $f_0^\ast$ of extinct patches in
the steady state. To leading order in $\delta$ we have $f_0^*(\delta) \sim 1-g_0(\delta)$. Thus the coefficient $c_1$ in Eq.~\eqref{eq:f0} is given by
\begin{equation}
        c_1 = \frac{I_1}{d_1 + m_{\rm c}}\Big(1 + \sum_{j=2}^{\infty} \frac{R_{1j}}{R_{11}} \Big)\, .
\end{equation}
Third, the coefficient $c_2$ determining the lowest order of an expansion of the number of individuals per patch, Eq.~(\ref{eq:mu}), in
powers of $\delta$ is given by the lowest-order term of Eq.~(\ref{eq:S1}). We find:
\begin{equation} 
c_2 = \frac{I_1}{m_{\rm c}}\,.
\end{equation}
Fourth, in order to find the coefficients $a_{11}$ and $a_{12}$ 
appearing in Eq.~\eqref{eq:Le1}, we insert Eqs.~\eqref{eq:partial_v_df2} and \eqref{eq:partial_v_dfdm} into Eq.~\eqref{eq:c11} for $\alpha = 1$. Using Eqs.~\eqref{eq:R} and \eqref{eq:L} yields
\begin{align}
\label{eq:a11_app}
a_{11} = {}&  L_{11} R_{11}
\bigg(d_1-m_{\rm c}\sum_{k=2}^\infty \frac{kR_{1k}}{R_{11}} \sum_{j=2}^k \frac{m_{\rm c}j}{d_j + m_{\rm c}j}\bigg)\,,\\
\label{eq:a12_app}
a_{12} = {}&-(d_1+m_{\rm c})  L_{11} R_{11}^2
\bigg(1+\sum_{k=2}^\infty \frac{R_{1k}}{R_{11}}\Big(1-k\sum_{j=2}^k \frac{m_{\rm c}}{b_{j-1}}\Big)\bigg) \,.
\end{align}
Finally, the coefficient $b_{11}$ in Eq.~\eqref{eq:slowH} is determined by inserting Eq.~\eqref{eq:partial_D} into Eq.~\eqref{eq:b11}, which results in
\begin{equation}\label{eq:b11_app}
b_{11} =  \sum_{i=2}^\infty (d_i + m_{\rm c} i) (L_{1i}-L_{1i-1})^2 R_{1i} - L_{11} \sum_{i=2}^\infty m_{\rm c}i (L_{1i}-L_{1i-1}) R_{1i} + d_1  L_{11}^2R_{11}\, .
\end{equation}

%****************************************************************************************************************
\section{Asymptotics for large values of $K$}
%****************************************************************************************************************
\label{app:C}
In this appendix we briefly demonstrate how to obtain the asymptotics
of the eigenvectors, of the coefficients  $a_{11}$, $a_{12}$, $b_{12}$, and of the critical
emigration rate $m_{\rm c}$  in the limit of $K\rightarrow \infty$
(see Eqs. (\ref{eq:mcexpand}) (\ref{eq:a11_a12_asy}), and (\ref{eq:asy}) in the main text). 
First, the results of this appendix demonstrate that the action
(\ref{eq:action2}) tends to $1/2$ as $K\rightarrow \infty$. Second,
the results obtained below shed light on the connection of our results to the work of
\citet{Lande1998}. 

In the limit of large carrying capacities $K$, the asymptotics of the coefficients $a_{11}$, $a_{12}$, $b_{12}$ 
is given by the asymptotics of $R_{11}$ which in turn is determined by the requirement that $\ve L_1\tr \ve R_1 = 1$. 
The first step consists of deriving the asymptotic form of $m_{\rm c}$, Eq.~(\ref{eq:mcexpand}), in the limit
of $K\rightarrow\infty$. The critical emigration rate $m_{\rm c}$ is given by Eq.~(\ref{eq:mc}):
\begin{equation}
\label{eq:mcC}
d_1 = m_{\rm c} \sum_{j=2}^\infty j \prod_{k=2}^j\frac{b_{k-1}}{d_k+m_{\rm c} k}\,.
\end{equation}
The product in this equation is estimated by
exponentiation, taking the continuum limit, and the resulting integral is evaluated in the 
saddle-point approximation, exactly as described by \citet{Doering2005}, see also \citep{Mehlig2007}.
Following \citet{Doering2005} we introduce the variable $x = i/K$ and define the functions $b(x)$ and $d(x)$ by
\begin{equation}
b_i = K\, b(x)\quad\mbox{and}\quad d_i = K\, d(x)\,.
\end{equation}
In the limit of $K\rightarrow \infty$, the term $m_{\rm c}k$ in the 
denominator in (\ref{eq:mcC}) is negligible compared to $d_k$.
The product is then estimated using
\begin{equation}
\prod_{k=1}^{j-1} \frac{b_k}{d_k} \sim {\rm e}^{K\int_0^y
{\rm d}x \log \rho(z) -\frac{1}{2}[\log\rho(0)+\log\rho(y)]}
= \frac{ {\rm e}^{-K \Phi(y)}}{\sqrt{\rho(0)\rho(y)}}
\end{equation}
with $y=j/K$, $\rho(x) = b(x)/d(x)$, and
\begin{equation}
\Phi(y) = -\int_0^y{\rm d} x\, \log \rho(x) = -\int_0^y
\, \frac{{\rm d}x\,r}{1+(r-1)x}\,.
\end{equation}
It follows that
\begin{equation}
 \sum_{j=2}^\infty d_1\frac{j}{d_j} \prod_{k=1}^{j-1} \frac{b_k}{d_k}
\sim K^2\int_0^\infty{\rm d}y\, \frac{d_1}{d(y)}
\frac{ {\rm e}^{-K \Phi(y)}}{\sqrt{\rho(0)\rho(y)}}\,.
\end{equation}
The integral is evaluated in  the saddle-point approximation.
The saddle point is $x_{\rm s} = 1$. With ${\rm d}^2\Phi/{\rm d}x^2=(r-1)/r$
at $x_{\rm s} = 1$ we have:
\begin{equation}
 \sum_{j=2}^\infty \frac{d_1}{d_j} \prod_{k=1}^{j-1} \frac{b_k}{d_k}
\sim K^2 \sqrt{\frac{2\pi r}{K (r-1)} } \frac{1}{Kr}
\frac{1}{\sqrt{r}} {\rm e}^{\Big[K\Big(1-\frac{\log r}{r-1}\Big)\Big]}\,,
\end{equation}
resulting in Eq.~(\ref{eq:mcexpand}).

The second step consists of determining the limiting form (\ref{eq:LlargeK}) of the elements of $\ve L_1$, Eq.~(\ref{eq:L}), in the limit of large values of $K$.
We now show that the double sum in Eq.~(\ref{eq:L}) grows exponentially, precisely cancelling the exponential
decrease of $m_{\rm c}$ as $K\rightarrow \infty$. The calculation is very closely related
to the evaluation of the time to extinction in a single-patch model by \citet{Doering2005}.
We use
\begin{equation}
\sum_{n=2}^{j-1} \,\,\sum_{k=n+1}^\infty \frac{k R_{1k}}{n R_{1n}} 
\sim \frac{K^2}{Kr} \sqrt{\frac{2\pi r}{K(r-1)}} {\rm e}^{\Big[K\Big(1-\frac{\log r}{r-1}\Big)\Big]} \sum_{n=2}^{j-1} 
\sqrt{\rho(0)} {\rm e}^{-n \log r}\,.
\end{equation}
Performing the geometric sum and inserting the result into Eq.~(\ref{eq:L}) we obtain
\begin{equation}
L_{1j} = 1-r^{-j}\,.
\end{equation}

The third step consists of estimating  $R_{11}$ which is determined by the requirement
\begin{equation}
\label{eq:LR}1=\ve L_1\tr \ve R_1 
\sim  R_{11}\Big(\frac{r-1}{r} + \sum_{j=2}^\infty (1-r^{-j})\frac{d_1}{d_j} \prod_{k=1}^{j-1} \frac{b_k}{d_k}\Big)\,.
\end{equation}
We proceed as before and find:
\begin{equation}
R_{11} \sim r\sqrt{\frac{K(r-1)}{2\pi }} {\rm e}^{\Big[-K\Big(1-\frac{\log r}{r-1}\Big)\Big]}\,.
\end{equation}

These results enable us to determine the coefficients 
 $a_{11}$, $a_{12}$, $b_{12}$ from Eqs.~(\ref{eq:a11_app}) to (\ref{eq:b11_app}).
In Eq.~(\ref{eq:a11_app}), the double sum is negligible compared to $d_1$ in 
the limit of $K\rightarrow \infty$. This implies
\begin{equation}\label{eq:a11_asy}
a_{11}\sim L_{11} R_{11} d_1 \sim \sqrt{\frac{K(r-1)^3}{2\pi }} {\rm e}^{\Big[-K\Big(1-\frac{\log r}{r-1}\Big)\Big]}\,.
\end{equation}
In Eq.~(\ref{eq:a12_app}), the sum over $j$ is negligible compare to unity. Evaluating the sum over $k$ in
the asymptotic limit, we find:
\begin{equation}
\label{eq:a12_asy}
a_{12}\sim -\sqrt{\frac{K(r-1)^3}{2\pi }} {\rm e}^{\Big[-K\Big(1-\frac{\log r}{r-1}\Big)\Big]}\,.
\end{equation}
Finally, Eq.~(\ref{eq:b11_app}) consists of three terms. The second term is negligible compared
to the other two. This implies:
\begin{equation}
b_{11} \sim \sqrt{\frac{K(r-1)^3}{2\pi }} {\rm e}^{\Big[-K\Big(1-\frac{\log r}{r-1}\Big)\Big]}\,.
\end{equation}
We see that the three coefficients have the same asymptotic dependence on $K$, apart from
the minus sign in Eq.~(\ref{eq:a12_asy}). It turns out that this asymptotic $K$-dependence
is exactly the inverse of the asymptotic dependence of the average time to extinction  $T_K$
upon $K$ for a single isolated patch with rates given by Eqs.~(\ref{eq:rates},\ref{eq:dr}) at carrying capacity $K$ (no migration). 
The exact expression for the average time to extinction starting with $i$ individuals
is \citep{NisbetGurney1982,vKa81,Doering2005}
\begin{equation}
\label{eq:Texact}
 T_i = \sum_{n=1}^{\infty} \frac{1}{d_n}\prod_{k=1}^{n-1} \frac{b_k}{d_k}
         + \sum_{n=1}^{i-1} \prod_{k=1}^{i} \frac{d_k}{b_k} 
\sum_{j=n+1}^{\infty} \frac{1}{d_j}\prod_{k=1}^{j-1}\frac{b_k}{d_k}\,.
\end{equation}
The corresponding asymptotic expression for $i=K$ and large values
of $K$ is given in Eq.~(19) of \citet{Doering2005}:
\begin{equation}
\label{eq:TKs}
T_K = \sqrt{\frac{2\pi}{K(r-1)^3}} {\rm e}^{\Big[K\Big(1-\frac{\log r}{r-1}\Big)\Big]}\,.
\end{equation}

Last but not least let us consider the probability $u_{1K}$ that
a single isolated patch grows from one individual to its carrying capacity $K$.
According to \citet{vKa81} one has:
\begin{equation}
	u_{1K} = \left({1 + \sum_{i=1}^{K-1} \prod_{j=1}^{i}\frac{ d_{j}}{ b_{j}}  }\right)^{-1} \,.
\end{equation}
This expression becomes independent of $K$ in the limit $K \rightarrow \infty$. Following the same procedure as above, we find
in this limit:
\begin{equation}
	u_{1K} \sim \frac{r-1}{r}\,.
\end{equation}
The corresponding probability to grow from $j$ individuals to $K$ is \citep{vKa81}:
\begin{equation}
	u_{jK} = \frac{\displaystyle 1 + \sum_{i=1}^{j-1} \prod_{k=1}^{i}\frac{ d_{k}}{ b_{k}}}{\displaystyle 1 + \sum_{i=1}^{K-1} \prod_{k=1}^{i}\frac{ d_{k}}{ b_{k}}  } \,,
\end{equation}
This expression tends to 
\begin{equation}
\label{eq:ujK}
	u_{jK} \sim 1 - r^{-j}
\end{equation}
as $K\rightarrow\infty$.

\end{appendices}

%****************************************************************************************************************
% Bibliography
%****************************************************************************************************************
%\newpage

%******************************************************************************************
%%%%%%%%%%%%%%%%%%%%%%%%%%%%%%%%%%%%%%%%%%%%%%%%%%%%%%%%%%%%%%%%%%%%%%%%%%%%%%%%%%%%%%%%%%%%
\newpage
\begin{center}
\begin{longtable}{lp{0.8\columnwidth}}
\caption{Symbols used in this article \label{tab:1}}\\
\toprule
\multicolumn{1}{c}{Symbol} & \multicolumn{1}{c}{Explanation} \\ 
\midrule
\endfirsthead
%%%%%%%%%%%%%%%%%%%%%%%%%%%%%%%%%%%%%%%%%
\multicolumn{2}{c}{{\tablename} \thetable{} (continued): Symbols used in this article} \\
\toprule
\multicolumn{1}{c}{Symbol} & \multicolumn{1}{c}{Explanation} \\ 
\midrule
\endhead
%%%%%%%%%%%%%%%%%%%%%%%%%%%%%%%%%%%%%%%%%%%
\\[-0.8em]
\multicolumn{2}{l}{{Continued on next page\ldots}} \\
\endfoot
%%%%%%%%%%%%%%%%%%%%%%%%%%%%%%%%%%%%%%%%%%%%
\bottomrule
\endlastfoot
%%%%%%%%%%%%%%%%%%%%%%%%%%%%%%%%%%%%%%%%%%%
$Q$			& 	Fraction of occupied patches, Eq.~\eqref{eq:LevinsModel} \\
$c$ 		& 	Colonisation rate, Eq.~\eqref{eq:LevinsModel}\\
$e$ 		& 	Extinction rate, Eq.~\eqref{eq:LevinsModel}\\
$N$ 		& 	Number of patches, Sec.~\ref{sec:model}  \\
$r$, $\mu$	&	Per-capita birth and death rates, Eqs.~(\ref{eq:rates},\ref{eq:dr}) \\
$K$			&	Patch carrying capacity, Eq.~(\ref{eq:dr}) \\
$m$     	&	Per capita emigration rate, Eq.~(\ref{eq:rates2})\\
$b_i$, $d_i$, $m_i$	& 	Birth-, death- and emigration rates for 
                            patch with $i$ individuals, Eqs.~(\ref{eq:rates},\ref{eq:dr},\ref{eq:rates2}) \\
$I$ 		& 	Patch immigration rate, Sec.~\ref{sec:model} \\
$M$ 		&	Number of migrants in common migration pool, Sec.~\ref{sec:model} \\
$\eta$	& 	Rate at which migrants leave common migration pool, Sec.~\ref{sec:model} \\
$\zeta$ 		& 	Rate of mortality during migration, Sec.~\ref{sec:model} \\
$\Lambda_k$	&	Rate for next event generated in patch $k$, Sec.~\ref{sec:nexp}\\
$\Lambda$ 	&	Rate for next event occurring in population, Sec.~\ref{sec:nexp}\\
$\ve n$ 	&   $\ve n = (n_1, n_2, \dotsc)^{\sf T}$ where $n_i$ is number of patches with $i$ individuals, Sec.~\ref{sec:mastereq} \\
$\rho(\ve n, t)$ & Probability of finding metapopulation in state $\ve n$ at time $t$, Eq.~\eqref{eq:mastereq} \\
$\mathbb{E}^{\pm}_j$ & Raising and lowering operators,  Eq.~(\ref{eq:rl}) \\
$\delta_{ij}$ & Kronecker delta: $\delta_{ij} = 1$ if $i=j$, and $\delta_{ij} = 0$ if $i \ne j$, Sec.~\ref{sec:mastereq}\\
$\ve f$		&	$\ve f = \ve n / N$: $f_i$ is fraction of patches with $i$ individuals, Sec.~\ref{sec:exp} \\
$\tilde\rho(\ve f, t)$ & $\tilde\rho(\ve f, t) = \rho(\ve n, t)$, Eq.~\eqref{eq:N1} (from Sec.~\ref{subs:detdyn} onwards, 
the tilde is dropped to simplify the notation) \\
$\ve v(\ve f)$ & Right-hand side of deterministic equation for $\ve f$, Eq.~\eqref{eq:det} \\
$S(\ve f)$ 	& 	Action, Eq.~\eqref{eq:ansatz} \\
$\ve p$ 	&	$\ve p = (p_1, p_2, \dotsc)^{\sf T}$: $p_i = \partial S / \partial f_i$, Sec.~\ref{sec:qssfinite} \\
$H(\ve f, \ve p)$ & Hamiltonian function, Eq.~\eqref{eq:H} \\
$\ma A$ & Jacobian matrix of the function $\ve v(\ve f)$:  $A_{ij} = \partial v_i/\partial f_j$, Sec.~\ref{sec:qssfinite} \\
$\ma D$ & Matrix with elements $D_{ij} = \partial^2H/\partial p_i\partial p_j$, Eq.~\eqref{eq:H2}\\
$\ma J$ & Jacobian matrix of the Hamiltonian dynamics, Eq.~\eqref{eq:J}\\
$\ma C$ 	& Covariance matrix (multiplied by $N$): $C_{ij} = N \operatorname{cov}[f_i, f_j]$, Eq.~\eqref{eq:C2} \\
$\sigma_j^2$ & Variance of $f_j$: for $j\geq 1$: $\sigma_j^2 = C_{jj} / N$, Sec.~\ref{sec:fluct}\\
$\ve f^*$ 	&	$\ve f$ at (quasi-)steady state, Eq.~\eqref{eq:fast} \\
$\delta \ve f$, $\delta \ve p$ & Disturbance of $\ve f$ and $\ve p$ away from $(\ve f^\ast,\ven 0)$, Eq.~\eqref{eq:bc0} \\
$I^*$ 		& 	Immigration rate at (quasi-)steady state, Eq.~\eqref{eq:sc} \\
$\ve p^*$ 	&	$\ve p$ at the fluctuational extinction point, Eq.~\eqref{eq:precursion} \\
$m_{\rm c}$	&   Critical emigration rate, Eq.~\eqref{eq:mcc}\\	
$\delta$ 	&   $\delta = (m - m_{\rm c})/m_{\rm c}$, Eq.~(\ref{eq:defdelta})\\
$\ma A^{(0)}$  & $\ma A$ evaluated at $\delta = 0$, Eq.~(\ref{eq:expand})\\
$\ma A^{(1)}$  & Matrix with elements $A^{(1)}_{ij} = \partial^2 v_i / \partial f_j \partial m$, evaluated at $\delta = 0$, Eq.~(\ref{eq:expand})\\
$A_{ijk}^{(2)}$  & $A^{(2)}_{ijk} = \partial^3 v_i / \partial f_j \partial f_k \partial m $, evaluated at $\delta = 0$, Eq.~(\ref{eq:expand})\\
$\lambda_{\alpha}$	& Eigenvalues of $\ma A$, ordered such that $\lambda_1 > \lambda_2 \geq \lambda_3 \geq \dotsi$, Eq.~\eqref{eq:lambda} \\
$\lambda_{\alpha}^{(0)}$	& Eigenvalues of $\ma A^{(0)}$, Eq.~\eqref{eq:lambda} \\
$\ve L_{\alpha}$, $\ve R_{\alpha}$ & Left and right eigenvectors of $\ma A^{{(0)}}$ with eigenvalues $\lambda_{\alpha}^{{(0)}}$, Eq.~\eqref{eq:lambda}\\%, and elements $L_{\alpha j}$\\
${\bf J}^{(0)}$	& ${\bf J}$ evaluated at $\delta=0$, Eq.~\eqref{eq:posevJ} \\
$\bm{\mathcal{R}}_{\alpha}$, $\bm{\mathcal{R}}_{\alpha}^{\prime}$  &  Right eigenvectors of $\ma J^{{(0)}}$, with eigenvalues $\lambda_{\alpha}^{{(0)}}$ and $-\lambda_{\alpha}^{{(0)}}$, Eq.~\eqref{eq:calR}  \\
$\bm{\mathcal{L}}_{\alpha}$, $\bm{\mathcal{L}}_{\alpha}^{\prime}$  &  Left eigenvectors of $\ma J^{{(0)}}$, with eigenvalues $\lambda_{\alpha}^{{(0)}}$ and $-\lambda_{\alpha}^{{(0)}}$, Eq.~\eqref{eq:calL} \\
$Q_{\alpha}$ & Slow ($\alpha=1$) and fast ($\alpha>1$) components of deterministic dynamics for $\ve f$, Eq.~(\ref{eq:expand2}) \\
$P_{\alpha}$ & Slow ($\alpha=1$) and fast ($\alpha>1$) momenta of the Hamiltonian dynamics, Eq.~\eqref{eq:QPalpha}\\
$B_{ijk}^{(1)}$ & $B^{(1)}_{ijk} = \partial D_{ij}/\partial f_k$, Sec.~\ref{sec:optpath}\\
$a_{11}$, $a_{12}$, $b_{11}$ & Coefficients in Eqs.~(\ref{eq:Le0},\ref{eq:slowH})\\
$Q_{1}^*$ & Slow variable $Q_1$ at (quasi-)steady state, Eqs.~(\ref{eq:fR},\ref{eq:s3}) \\
$P_{1}^*$ & Slow variable $P_1$ at fluctuational extinction point, Eqs.~(\ref{eq:s2},\ref{eq:pR})\\
%$c_1$, $c_2$	& Coefficients involved in the steady state of the deterministic dynamics (for small $\delta$)\\
$T_{\rm ext}$ 	& Expected time to extinction of the metapopulation, Eq.~\eqref{eq:expNS}\\
$T_K$	&	Expected time to extinction of a single patch, Sec.~\ref{sec:text}\\
$u_{jK}$	&	Probability that patch with $j$ individuals reaches $K$ individuals before becoming extinct, Eq.~\eqref{eq:cLande}\\
$n$		&	Total number of individuals in metapopulation, Sec.~\ref{app:D} \\
$M_n(i_1,\dotsc,i_N)$	&	Multinomial distribution of $i_1,\dotsc,i_N$, with $\sum_{k=1}^{N} i_k = n$, Eq.~\eqref{eq:mn} \\
$B_n$, $D_n$	& 	Global birth and death rates in the limit $m \to \infty$, Eqs.~(\ref{eq:Bn}, \ref{eq:Dn}) \\
$K_{\rm tot}$ & Global carrying capacity in the limit $m \to \infty$, Sec.~\ref{app:D}\\
%$g_0(I_1)$, $g_1(I_1)$	& Functions involved in the expansion of $I$, Appendix \ref{app:B}\\
%$b(x)$, $d(x)$	& $b(x) = b_i$, $d(x) = d_i$, where $x=i/K$, Appendix \ref{app:C}\\
%$\Phi(x)$	& Function involved in the calculation of the asymptotic value of $m_{\rm c}$ for large $K$, Appendix \ref{app:C}\\
%\bottomrule
\end{longtable}
%\end{table}
\end{center}

%%%%%%%%%%%%%%%%%%%%%%%%%%%%%%%%%%%%%%%%%%%%%%%%%%%%%%%%%%%%%%%%%%%%%%%%%%%%%%%%%%%%%%%%%%%%

\end{document}